\documentclass[aps ,twocolumn, secnumarabic ,amssymb, amsmath,nobibnotes, aps, prl, floatfix]{revtex4-1}
\usepackage{docs}
\usepackage{bm}
\usepackage[pdftex,colorlinks=true,linkcolor=blue]{hyperref}%
\expandafter\ifx\csname package@font\endcsname\relax\else
 \expandafter\expandafter
 \expandafter\usepackage
 \expandafter\expandafter
 \expandafter{\csname package@font\endcsname}%
\fi

\usepackage{graphicx}
\usepackage{url} 

\DeclareMathOperator\erf{erf}
\numberwithin{equation}{section}

\newcommand{\aap}{Astronomy \& Astrophysics}
\newcommand{\icarus}{Icarus}
\newcommand{\mnras}{Monthly Notices of the Royal Astronomical Society}
\newcommand{\jqsrt}{J. Quant. Spectr. Rad. Trans.}
\usepackage{color}

\begin{document}

\title{Scale Dependence of Multiplier Distributions for Particle Concentration, Enstrophy and Dissipation in the Inertial Range of Homogeneous Turbulence}
\author{Thomas~Hartlep}
\affiliation{Bay Area Environmental Research Institute, Petaluma, CA 94952, USA}
\altaffiliation[Also, ]{NASA Ames Research Center, Moffett Field, CA 94035, USA}
\email{thomas.hartlep@nasa.gov}
\author{Jeffrey~N.~Cuzzi}
\affiliation{NASA Ames Research Center, Moffett Field, CA 94035, USA}
\author{Brian~Weston}
\affiliation{University of California Davis, Mechanical \& Aerospace Engineering, Davis, CA 95616, USA}

\date{12 December 2016 (Submitted to {\it Physical Review E}), 28 February 2017 (Accepted)}

\begin{abstract}

Turbulent flows preferentially concentrate inertial particles depending on their stopping time or Stokes number, which can lead to significant spatial variations in the particle concentration.
Cascade models are one way to describe this process in statistical terms.
Here, we use a direct numerical simulation (DNS) dataset of homogeneous, isotropic turbulence to determine probability distribution functions (PDFs) for cascade \textit{multipliers}, which determine the ratio by which a property is partitioned into sub-volumes as an eddy is envisioned to decay into smaller eddies.
We present a technique for correcting effects of small particle numbers in the statistics.
We determine multiplier PDFs for particle number, flow dissipation, and enstrophy, all of which are shown to be scale dependent.
However, the particle multiplier PDFs collapse when scaled with an appropriately defined \textit{local} Stokes number.
As anticipated from earlier works, dissipation and enstrophy multiplier PDFs reach an asymptote for sufficiently small spatial scales.
From the DNS measurements, we derive a cascade model that is used it to make predictions for the radial distribution function (RDF) for arbitrarily high Reynolds numbers, $Re$, finding good agreement with the asymptotic, infinite $Re$ inertial range theory of Zaichik \& Alipchenkov [New Journal of Physics 11, 103018 (2009)].
We discuss implications of these results for the statistical modeling of the turbulent clustering process in the inertial range for high Reynolds numbers inaccessible to numerical simulations.

\end{abstract}


\maketitle


\section{Background and Introduction}
\label{Section:Introduction}

 Clustering of inertial (finite-stopping-time) particles into dense zones in fluid turbulence has applications in many fields \citep[for a general review see][]{2009AnRFM..41..375T}. A number of recent papers have focussed on understanding the basic mechanisms responsible for this effect; several of these \citep{ZaichikAlipchenkov2009, BraggCollins2014I, BraggCollins2014II, GustavssonMehlig2014} provide thorough reviews and comparisons of previous studies dating back to the early work of \citet{Maxey1987} and \citet{SquiresEaton1991} which we will only sketch briefly.  The early work emphasized the role of centrifugation of finite-inertia particles out of vortical structures in turbulence. More recent evidence that clustering arises even in random, irrotational flows suggests that, while vorticity still plays a role, the dominant role is played by so-called ``history effects", in which inertial particle velocity dispersions at any location carry a memory of particle encounters with more remote flow regimes which have larger characteristic velocity differences \citep{PanPadoan2010, 2013ApJ...776...12P, Braggetal2015JFM}. These history effects lead to spatial gradients in particle random relative velocities, and these gradients in turn generate systematic flows or currents which can outweigh dispersive effects and produce zones of highly variable particle concentration \citep{ZaichikAlipchenkov2003, ZaichikAlipchenkov2009, BraggCollins2014I, BraggCollins2014II}. 

To date, by far the most attention regarding particle clustering in turbulence  has been devoted to very small spatial scales $r < \eta$ or even $r \ll \eta$, where $\eta$ is the Kolmogorov scale, partly because it is on these scales that particle collisions occur and partly because numerical simulations to date have produced only very limited inertial ranges, at best \citep[see however][]{2010JFM...646..527B, Irelandetal2015}. Theories by \citet{ZaichikAlipchenkov2003} {\it et seq.}, and \citet{PanPadoan2010} {\it et seq.} have been shown to be promising in explaining the cause of particle clustering in terms of history effects, with helpful contributions from the traditional local centrifugation mechanism \citep{ZaichikAlipchenkov2009, BraggCollins2014II, Braggetal2015JFM, Irelandetal2015}. A thorough review of the effects of clustering and relative velocity effects on particle collisions, emphasizing the astronomical literature, can be found in \citet{PanPadoan2014, PanPadoan2015}.

Our focus is on clustering at larger scales in the {\it inertial range} $\eta < r < L$, where $L$ is the integral scale. Inertial range clustering  has important applications for remote sensing of terrestrial clouds \citep{Shawetal2002}, the formation of primitive planetesimals (asteroids and comets) in the early solar nebula \citep{2008ApJ...687.1432C, 2010Icar..208..518C, Chambers2010,0004-637X-740-1-6, 2014ApJ...797...59H, Johansenetal2015}, and even the structure of the interstellar medium \citep{2015arXiv151002477H}. While little studied in the context of particle clustering, inertial range scaling is known to have different  properties than seen in the dissipation range $r < \eta$ \citep{2007PhRvL..98h4502B, Braggetal2015PRE}. Only limited predictions have been made of its scaling properties at very high Reynolds number $Re$ \citep{ZaichikAlipchenkov2003,ZaichikAlipchenkov2009}. 

In the inertial range, so-called {\it cascade models} which reproduce the statistics of fluid behavior, even if not realistic flow structures, may be valuable for modeling high Reynolds number ($Re$) regimes too demanding for direct numerical simulations.
 Their application is quite general~\citep[][see~\cite{2007PhRvE..75e6305H} for more references]{1987PhRvL..59.1424M, 1991JFM...224..429M, 1989PhRvA..40.5284C, 1992PhRvL..68.2762C, 1994ActaMechSup4..113S, 1995JSP....78..311S}. We and others have used cascades to model particle clustering in turbulence in the astronomical applications mentioned above. 
 
\begin{figure*}
\begin{center}
\includegraphics[width=0.9\linewidth]{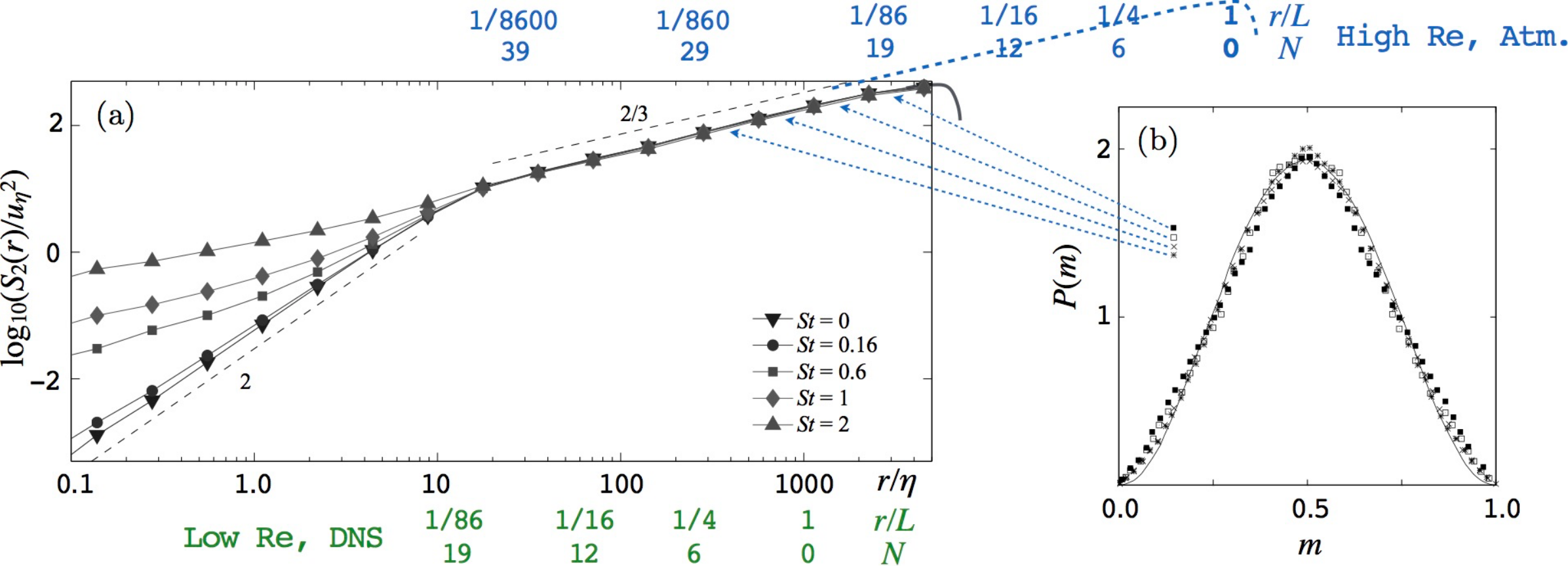}
\caption{(a) The second order structure function for particle velocity for the DNS data we analyze. Statistics are obtained over the trajectories of particles of different $St$, from figure~2 of~\citet{2010JFM...646..527B}. In particular circles and downward facing triangles refer to tracer-like particles that more faithfully follow the fluid flow. Dotted power laws labeled ``2" and ``2/3" are the theoretical expectations for the flow velocity structure function in the viscous range and the inertial range, respectively. The bottom axis is labeled as in~\citet{2010JFM...646..527B} by $r/\eta$, and also by us (in green) with our estimated values for $r/L$, where $L$ is the integral scale, and also by the cascade level $N$ that gives equidimensional volumes of side $r$. Extending out the top of the plot is an offset blue short-dashed line indicating the expected structure function for the high-$Re$ atmospheric flow of~\citet{1987NuPhS...2...49M}, analyzed by~\citet{1989PhRvA..40.5284C} and~\citet{1995JSP....78..311S}, with integral scale which we estimate as $L \sim 2 \times 10^5 \eta$. We have attempted to place a corresponding scale of $r/L$ and $N$ on the top axis (in blue).
(b) Scale-independent multiplier PDFs for dissipation $\epsilon$ as determined by~\citet{1995JSP....78..311S} in the atmospheric boundary layer, connected to the left panel by dotted arrows indicating where those measurements lie on the structure function (300-3000 $\eta$, well away from both $L$ and $\eta$). It is our expectation that the scale-independent $\beta$-distribution with $\beta \sim 3$ (see equation~\ref{Eqn:BetaPDF}) observed by~\citet{1995JSP....78..311S} (smooth curve in right panel) continues at least to the start of the viscous range at about 30$\eta$. Our analyses of the DNS dataset are binned on scales between $r = 12\eta-512\eta$, corresponding to $r \sim L/2$ to $L/86$ on the lower scale as discussed in in section~\ref{Section:Introduction}.
\label{fig:inertialrange}}
\end{center}
\end{figure*}

Cascade models operate by simply applying a partition function or \textit{multiplier} $0 \leq m \leq 1$ to any property $\cal P$ in some given volume of the flow, thus determining the ratio by which the property (dissipation, particle density, etc.) is partitioned into sub-volumes as an eddy is envisioned to decay into smaller eddies. The most common treatment is a binary cascade, in which $\cal P$ is partitioned into two equal subvolumes; however the approach can be applied to arbitrary numbers of subvolumes~\citep{1994ActaMechSup4..113S}. 
The binary cascade operates on each volume of space, partitioning $\cal P$ into two equal subvolumes by multipliers $m$ and $1-m$, with the multiplier $m$ at each bifurcation drawn from a probability distribution function (PDF) of multipliers $P(m)$.
If $P(m)=\delta(m-0.5)$, where $\delta$ is the delta function, the cascade has no effect because the property $\cal P$ is evenly divided, and remains constant per unit volume.
On the other hand, broad $P(m)$ functions generate highly \textit{intermittent} spatial distributions in which $\cal P$ has a wide range of values, fluctuating dramatically on small scales such as seen in dissipation~\citep{1991JFM...224..429M, 1995JSP....78..311S} (figure~\ref{fig:inertialrange}b). 

The \textit{dissipation range}, a range of small scales approaching the Kolmogorov scale $\eta$, is found where $r < 20-30\eta$~\citep{1972fct..book.....T,1995tlan.book.....F}; in this range, where viscosity is important, the equations of motion are no longer completely scale-free, and fluid scaling properties differ from those in the inertial range. The properties of particle clustering {\it do} seem to be scale independent in this region, however \citep{2007PhRvL..98h4502B, Braggetal2015PRE}, and one expects this regime to be flow-independent for high $Re$. 
There is also a range of \textit{large} scales near the integral scale $L$, over which deviation from scale invariance surely occurs, but this range has not been well studied and is surely flow-dependent.
The application to planetesimal formation has become focussed on particle concentration at scales much larger than the Kolmogorov scale~\citep{2008ApJ...687.1432C,2010Icar..208..518C} because large clumps are needed for sufficiently rapid gravitational collapse.
In previous particle clustering cascade models,~\citet{2007PhRvE..75e6305H} determined the multiplier PDFs for particle concentration and fluid enstrophy at \textit{small} spatial scales, not too far from $\eta$ (to obtain better statistics), and applied them across all scales ranging up to the integral scale 
 (see section \ref{Subsection:Discussion:Particles} for more discussion). Realizing the risks in this, they performed tests which seemed to validate the approach. However, discrepancies between~\citet{2007PhRvE..75e6305H} and~\citet{0004-637X-740-1-6} at the low probabilities of interest for the planetesimal problem~\citep{2010Icar..208..518C} have led us to explore the scale dependence of $P(m)$ in more detail, in order to improve the fidelity of the cascade models. 

It is worth noting at this point that it is not a requirement of cascade models that the PDFs be scale-independent; it is merely the first and most obvious assumption.
In this paper we present evidence that the multiplier PDFs for particle concentration are scale-\textit{dependent} and present simple guidelines for how this scale-dependence can be included in cascades.
Multiplier PDFs can also be \textit{conditioned} on local properties~\citep{1995JSP....78..311S}, and indeed were treated this way in our previous work to allow for particle mass-loading on the process~\citep{2007PhRvE..75e6305H}.
Scale-dependence \textit{per se} is, however, a different effect than local conditioning, and in this paper we do not address local conditioning.

\begin{figure}
\begin{center}
\hspace*{-0.0cm}\includegraphics[width=1.00\linewidth, angle=0.0]{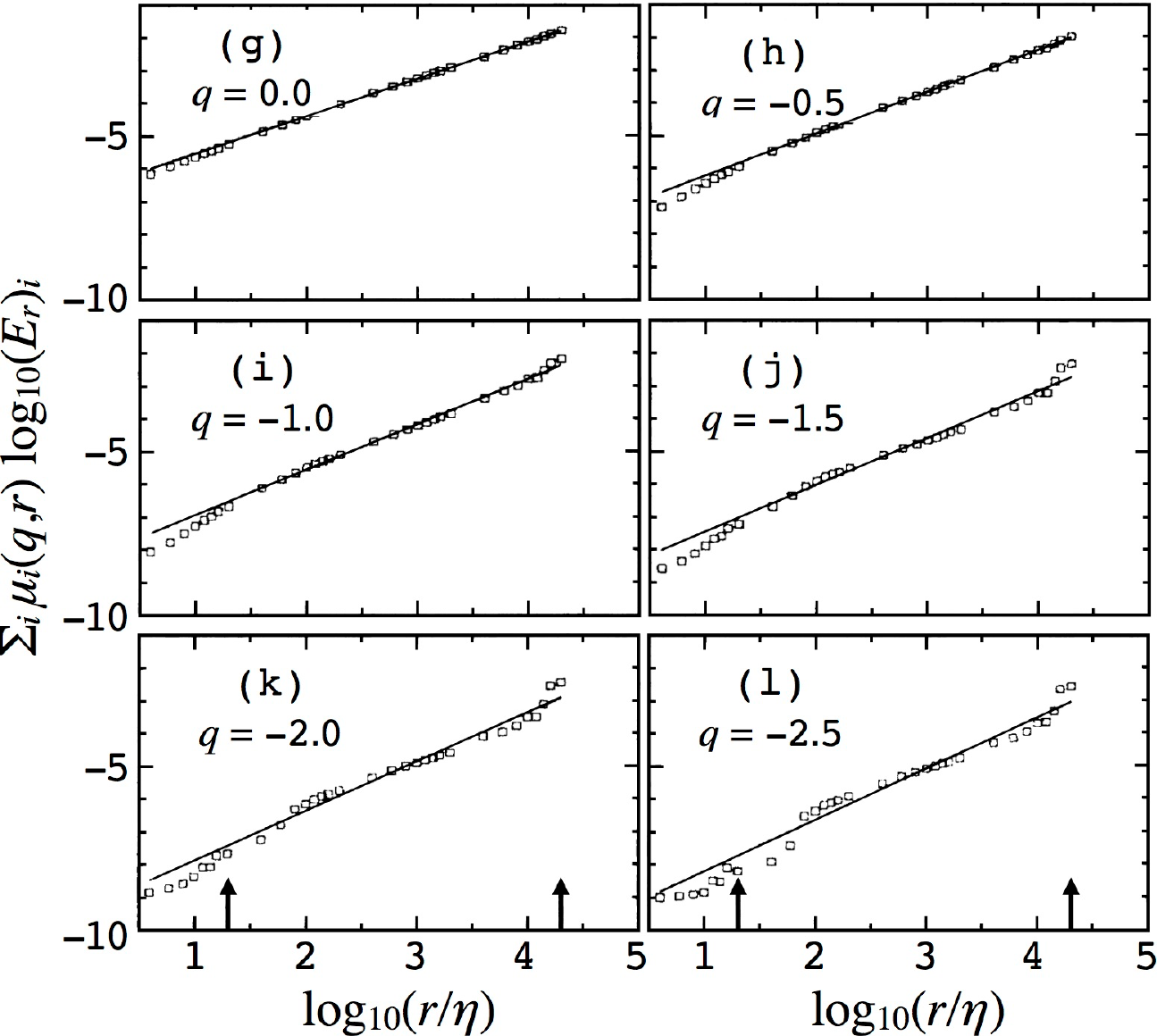}
\caption{Plot of a family of normalized $q$-th moments of the dissipation $E_r$ in an atmospheric boundary layer, as averaged over binning lengthscale $r$, plotted against $r/\eta$ (taken from figure~5b of~\citet{1989PhRvA..40.5284C}). The quantity $\mu_i(q,l) \equiv (E_r)_i^q/\Sigma_j(E_r)_j^q$, where $(E_r)_i$ is the dissipation in the $i$th bin of size $r$. Smaller $|q|$ values suppress the effect of strongly anomalous regions, while large negative values of $q$ select for regions of anomalously low turbulent dissipation. Abrupt changes in the slope of these plots (most obvious for large $|q|$) might indicate departure from the true scale-free inertial range, both in the dissipation range at $< 20-30\eta$, and at very large scales where vortex stretching has yet to become effective. Vertical arrows on horizontal axis are the authors~\citep{1989PhRvA..40.5284C} estimate of the inertial scaling range, but the scaling range is narrower for larger $|q|$. 
\label{fig:chhabra}}
\end{center}
\end{figure}

Before describing our own work, we review some experimental results on high-Reynolds number atmospheric boundary layer turbulence, which provide a useful background in scale invariance and complement the more typical, but lower-$Re$, numerical simulations. 
Studies of the properties of turbulence in atmospheric boundary layer flows have been conducted by~\citet{Kholmyanskiy:1973tg, 1975JFM....71..417V} and~\citet{1987NuPhS...2...49M}; further analysis of the~\citet{1987NuPhS...2...49M} data was done by~\citet{1989PhRvA..40.5284C}.\footnote{These results were cited and reanalyzed by~\citet{1995JSP....78..311S}, however the reference to the basic data given by~\citet{1995JSP....78..311S} is confused with an interpretive article by~\citet{1989PhRvL..62.1327C}, who themselves cite~\citet{1987PhRvL..59.1424M} and~\citet[at the time unpublished]{1989PhRvA..40.5284C} for discussions and analysis of the basic data.} The best reference for the basic experimental data is~\citet[see their Table 1]{1987NuPhS...2...49M}, who conducted an experiment on boundary layer turbulence using a sensor mounted 2~m above the flat roof of a four story building. 
The Reynolds number for the flow is calculated using the free stream velocity $U$ = 6 m/s and the height $h$ = 2~m of the sensor above the roof: $Re = Uh/\nu = 8 \times 10^5$ where the kinematic viscosity is $1.5\times10^{-5}~$m$^2$/sec, consistent with tabulated values in~\citet{1987NuPhS...2...49M} of the Taylor scale Reynolds number $Re_{\lambda}$, and its characteristic lengthscale $\lambda$ and velocity $u'$. \citet{1987NuPhS...2...49M} give the Kolmogorov scale as $\eta = 7 \times 10^{-4}$~m (we retain their preferred units). Analyses of flow structures by~\citet{1989PhRvA..40.5284C} (their figure 5 and our figure \ref{fig:chhabra}) show fairly well-behaved power law scaling of dissipation for weightings which suppress regions that are strongly anomalous (panels g and h), i.e. strongly differing from the mean, to almost $r \sim 1.8\times 10^4~\eta$ = 12.6~m $\gg h$, suggesting an extensive inertial range. The large or integral scale $L$, which contains the energy in this flow, is thus apparently much larger than the vertical distance of the sensor from the boundary ($h = 2$~m) and plausibly the same as the \textit{longitudinal} integral scale given as $L$ = 180~m~\citep{1987NuPhS...2...49M}, see also~\citet{Hunt:2000kb}, and thus $L/\eta \sim 2 \times 10^5$. 
However, when the role of strongly anomalous regions is emphasized (panels i through l of figure~\ref{fig:chhabra}) the scalable inertial range contracts.

In cascade applications, it may be more meaningful to assess the scale dependence of $P(m)$ at large scales not in terms of multiples of $\eta$ as in~\citet{1995JSP....78..311S} and most other work~\citep[e.g.,][]{2007PhRvL..98h4502B}, but in terms of fractions of $L$ which more closely connects to causality and energy flow. We will also express scale fractions $r/L$ in terms of cascade bifurcation \textit{levels} $N$ needed to achieve cubes $r$ on a side: 
\begin{equation}
 r/L = 2^{-N/3}.
 \label{Eqn:r_over_L}
\end{equation}

For example, \citet{1995JSP....78..311S} compared multiplier PDFs for dissipation in the atmospheric boundary layer over a range of scales (see figure~\ref{fig:inertialrange}b) and showed that $P(m)$ is highly scale independent over a wide range of scales: $372\eta-3072\eta$, or $372\eta$ to $L/86$ ~\citep[see also][]{1992PhRvL..68.2762C}. Dissipation depends on higher-order moments of the velocity gradients, so we are  drawn for guidance to the behavior seen in the higher-order moments (larger $|q|$) in figure~\ref{fig:chhabra}~\citep{1989PhRvA..40.5284C}. The results of~\citet{1995JSP....78..311S} are consistent with the \textit{generally} power law behavior seen for $25\eta - 10000\eta$ (roughly $25\eta$ to $L/26$) in figure~\ref{fig:chhabra}.  That is, one might infer from where the plots in~\citet{1989PhRvA..40.5284C} deviate from power law behavior, that the scale-free behavior demonstrated by~\citet{1995JSP....78..311S} (figure~\ref{fig:inertialrange}b) might carry on to larger sizes than they actually presented, possibly until $r \sim L/26$ or $10000\eta$, but deviate noticeably for scales larger  than $r \sim L/15$ (and at the smaller end below $25\eta$).
Moreover, we can conclude from these comparisons that the scale-free behavior seen by~\citet{1995JSP....78..311S} was safely out of the viscous range, and continued through the inertial range at scales up to $L/86 < L/10$.  

The goal of this paper is to use DNS data to derive probability distribution functions for cascade multipliers and construct a cascade model that can be used for modeling higher $Re$-number flows not accessible to direct numerical simulations.
The paper is organized as follows: section~\ref{Section:Dataset} describes the DNS dataset used in this study; section~\ref{Section:Analysis} describes the data analysis including a novel technique for correcting the effects of small particle number statistics, and presents results for the multiplier PDFs for particle concentration, dissipation and enstrophy; section~\ref{Section:NewCascade} presents predictions of the new cascade model and comparison with DNS data at two different $Re$; and section~\ref{Section:Discussion} discusses the results and their implications. A summary and conclusions are given in section~\ref{Section:Conclusions}.

\section{Dataset}
\label{Section:Dataset}

In this paper, we use data from the direct numerical simulations of~\citet{2010JFM...646..527B}; see also~\citet{2008PhRvL.100y4504A} and~\cite{2009PhRvE..80f6318B}.
The simulation computes forced, homogenous and isotropic turbulence in an incompressible fluid, and the dynamics of inertial particles suspended in the flow.
The fluid flow is solved on a $2048^3$ Cartesian grid with a grid spacing that is approximately the Kolmogorov length scale $\eta \approx \Delta x = \Delta y = \Delta z$.
Tracer and inertial particles are introduced into the flow and their trajectories are tracked. 
Particles are considered point particles and are dragged with the flow by viscous forces only; there is no back-reaction on the flow.
Particles of different Stokes numbers $St \equiv \tau_s / \tau_\eta$ are considered, where $\tau_s$ is the aerodynamic stopping time of the particle ($\tau_s = 0$ for tracers) and $\tau_\eta$ is the Kolmogorov time.

Figure~\ref{fig:inertialrange}a shows the second order structure function for particle velocity for this numerical flow~\citep[taken along trajectories of different $St$ particles, from figure~2 of][]{2010JFM...645..497B}, as a function of normalized scale $r/\eta$. While the structure function seems to show an inertial range to several thousand $\eta$, in reality the integral eddy scale for this simulation seems to be about half the computational box size, $L \sim 1024\eta$, and for this flow $Re_{\lambda} \approx 4(L/\eta)^{2/3} \sim$ 400~\citep[table 1 of][]{2010JFM...645..497B}.
In blue-green below the horizontal axis we indicate the corresponding values for $r/L$, and the corresponding cascade level $N$.
The blue dashed line indicates the expected inertial range for the atmospheric flow of~\citet{1987NuPhS...2...49M}, with corresponding values of $r/L$ and $N$ also indicated in blue above the figure. Note that the range where $P(m)$ for dissipation was observed to be scale independent by~\citet{1995JSP....78..311S} corresponds roughly to the scale range $L/860 - L/86$, well below the expected integral scale for that flow and well above the viscous subrange.

Data from this simulation are available publicly online~\citep{RM-2007-GRAD-2048}, and we have downloaded and analyzed all of the publicly available data in the present work.
This data consists of the entire flow field sampled at 13 instances in time, and particle trajectories sampled at 4,720 equidistant times, both covering about 6 large-eddy time scales $\tau_L$.
All flow components and their first derivatives are available at the particle locations.
In total there are $3 \times 64$ files of particle trajectories each containing 3,184 particles (a total of $N_p \approx 600\textrm{k}$ particles) for each $St = 0, 0.16, 0.6$ and $1.0$, and 64 files containing 3,184 particles each (i.e., a total of $N_p \approx 200\textrm{k}$ particles) for each $St = 2, 3, 5, 10, 20, 30, 40, 50,$ and $70$.


\section{Analysis}
\label{Section:Analysis}\

Determining concentration multipliers amounts to counting particles in cubic sub-volumes of size $r^3$ and calculating the fraction of particles falling in each half of the sampling box.
We bisect each cube in all three orthogonal directions $x$, $y$, and $z$ each yielding 2 multiplier values totaling 6 multiplier values for $r^3$ cube.
The available trajectory data is highly resolved in time (4,720 instances of time over approximately 6 large eddy times $\tau_L$), much more than what is needed for this analysis.
The number of snapshots required for good statistics depends on the scale of interest since structures at large scale evolve more slowly than structures at small scale (and contain more particles), and therefore can be sampled less often.
We choose to sample the particle data with a temporal spacing of $\tau_\textrm{sample} \approx 0.55~\tau_r$ where $\tau_r$ is the characteristic eddy life time at spatial scale $r$ estimated using Kolomogorov 1941 arguments as $\tau_r = \tau_L (r/L)^{2/3}$.
For the box sizes considered, $512\eta$, $256\eta$, $128\eta$, $64\eta$, $45\eta$, $32\eta$, $24\eta$, $16\eta$ and $12\eta$, this results in sampling intervals of $0.34\tau_L$, $0.22\tau_L$, $0.14\tau_L$, $0.086\tau_L$, $0.067\tau_L$, $0.055\tau_L$, $0.044\tau_L$, $0.034\tau_L$ and $0.028\tau_L$, respectively.
We populate the sample volume using the positions of all particles from the high-resolution trajectory files at these various discrete times.

\subsection{Tracer particles}
\label{Subsection:Analysis:Tracers}

\begin{figure*}
   \centering
   \hspace*{-0.0cm}\vspace*{0.0cm}\includegraphics[width=0.9\linewidth]{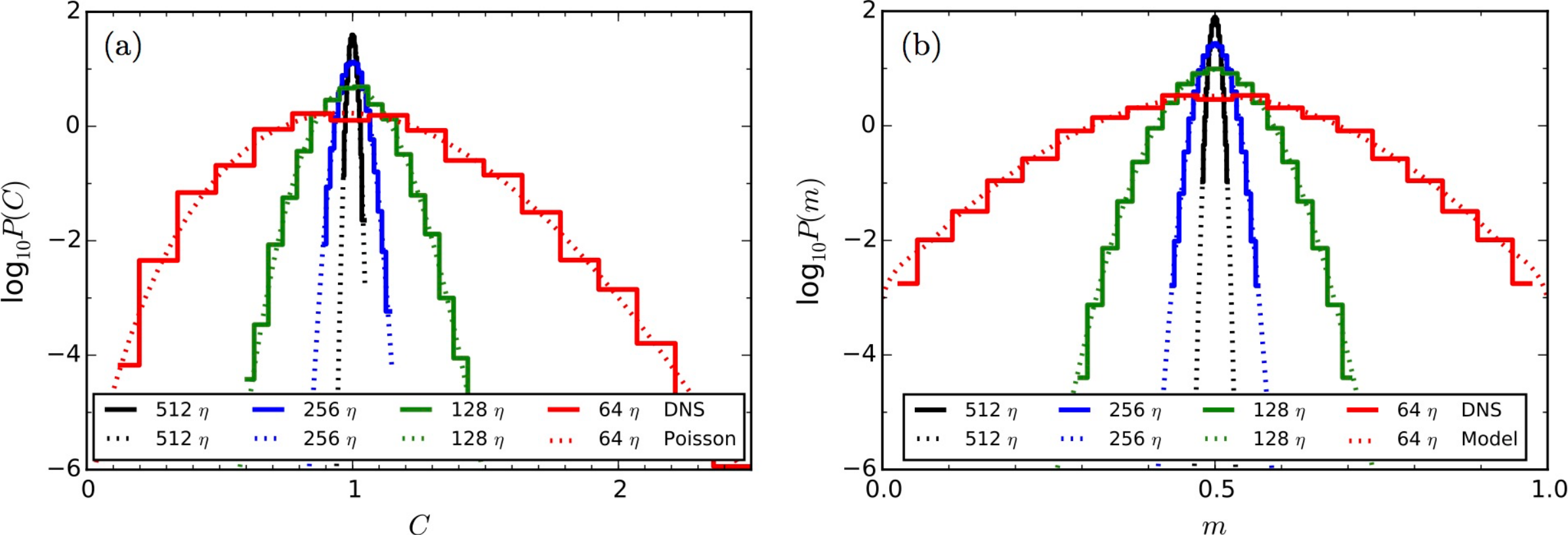}
   \caption{(a) Probability distribution functions (PDFs) of the particle concentration itself for tracer particles ($St=0$) from the simulation dataset, compared to Poisson distributions, for box sizes of $r=512\eta$, $256\eta$, $128\eta$, and $64\eta$. (b) PDFs of the particle concentration multipliers for $St=0$ particles computed from the simulation dataset, and the analytical Poisson/Binomial model from equation~(\ref{Eqn:PoissonBinomial}).
   \label{Fig:TracersConcentrationAndMultiplierPDF}}
\end{figure*}

First, we will look at the \textit{tracer} particles (Stokes number $St=0$) which follow the flow exactly.
Particles are initially homogeneously distributed, and since the flow is incompressible, will stay that way on the average, with particle multiplier distributions given by a delta function at $m=0.5$.
However, the observed distributions will only approach this behavior for very large number of particles; otherwise, effects of small number statistics complicate matters.
The simulation dataset at hand has $N_p \approx 6\times10^5$ particles with $St=0$, and this is small enough to cause significant deviations from ideal behavior at small scales. 

For $St=0$ particles, we can quantify these small-number effects analytically.
First, the probability of finding $n$ particles in a given box of size $r$ is governed by a Poisson distribution
\begin{equation}
   \label{Eqn:Poisson}
   P_P(n;\bar{n}) = \frac{\bar{n}^n e^{-\bar{n}}}{n!}
\end{equation}
with the expectation value $E(n) = \bar{n}(r)$, which is here the average number of particles in a box of size $r$, that is $\bar{n}(r) = N_p r^3 / {\cal L}^3$, and where $\cal L$ is the size of the simulation domain.
This probability is also the probability of finding a concentration $C=n/{\bar n}$.
A comparison of the observed probability distribution functions with equation~(\ref{Eqn:Poisson}) is shown in figure~\ref{Fig:TracersConcentrationAndMultiplierPDF}a.

Then, for a given box with exactly $n$ particles, consider the particles one at a time and ask if they fall into one half of the box, say the left half, or the other, right half.
For tracer particles, the probability to fall into the left side is the same as falling into the right side of the box, i.e. $p_\textrm{left}=p_\textrm{right}=p=0.5$, and the probability of having exactly $k$ particles fall into one side of the box is then given by a binomial distribution with the probability
\begin{equation}
   \label{Eqn:Binomial}
   P_B(k;n,p) = {n \choose k} p^n (1-p)^{n-k}.
\end{equation}
This is then also the probability of finding a multiplier $m=k/n$ in a box of $n$ particles. 

By combining the probabilities~(\ref{Eqn:Poisson}) and~(\ref{Eqn:Binomial}), we see that the probability of finding a multiplier $m$ in the entire simulation domain is
\begin{equation}
\label{Eqn:PoissonBinomial}
P(m;\bar{n},p) = \sum_n P_P(n;\bar{n}) P_B(k=mn;n,p).
\end{equation}
A comparison of this analytical relation with the multiplier PDFs computed from the tracer particle trajectories in the simulation is shown in figure~\ref{Fig:TracersConcentrationAndMultiplierPDF}b.
It shows that equation~(\ref{Eqn:PoissonBinomial}) models the observed distributions very accurately, and also that the distributions become rather wide at small scales even though the underlying probability distributions are delta functions at $m=0.5$.
This effect is a kind of ``false intermittency'' due to small-number statistics alone.

\subsection{Particle multipliers for non-zero Stokes numbers}
\label{Subsection:Analysis:NonZeroStokesNumbers}

\subsubsection{Correcting for finite particle numbers}
\label{Subsubsection:Correcting}

As we have seen in the previous section, the number of particles in the dataset is small enough to significantly affect the observed multiplier distributions.
In the following we will describe how we can account for these effects and estimate what the underlying PDFs would be, given infinite particle numbers.
The goal is to separate the finite-particle number effects, which may be important in many applications, from the effects of the turbulent concentration process.
The finite-particle-number effects in a specific application can always be added back into our model later (see, e.g., section~\ref{Subsection:ComparisonCascadeWithDNS}). 

\begin{figure}
   \centering
   \hspace*{-0.0cm}\includegraphics[width=0.95\linewidth]{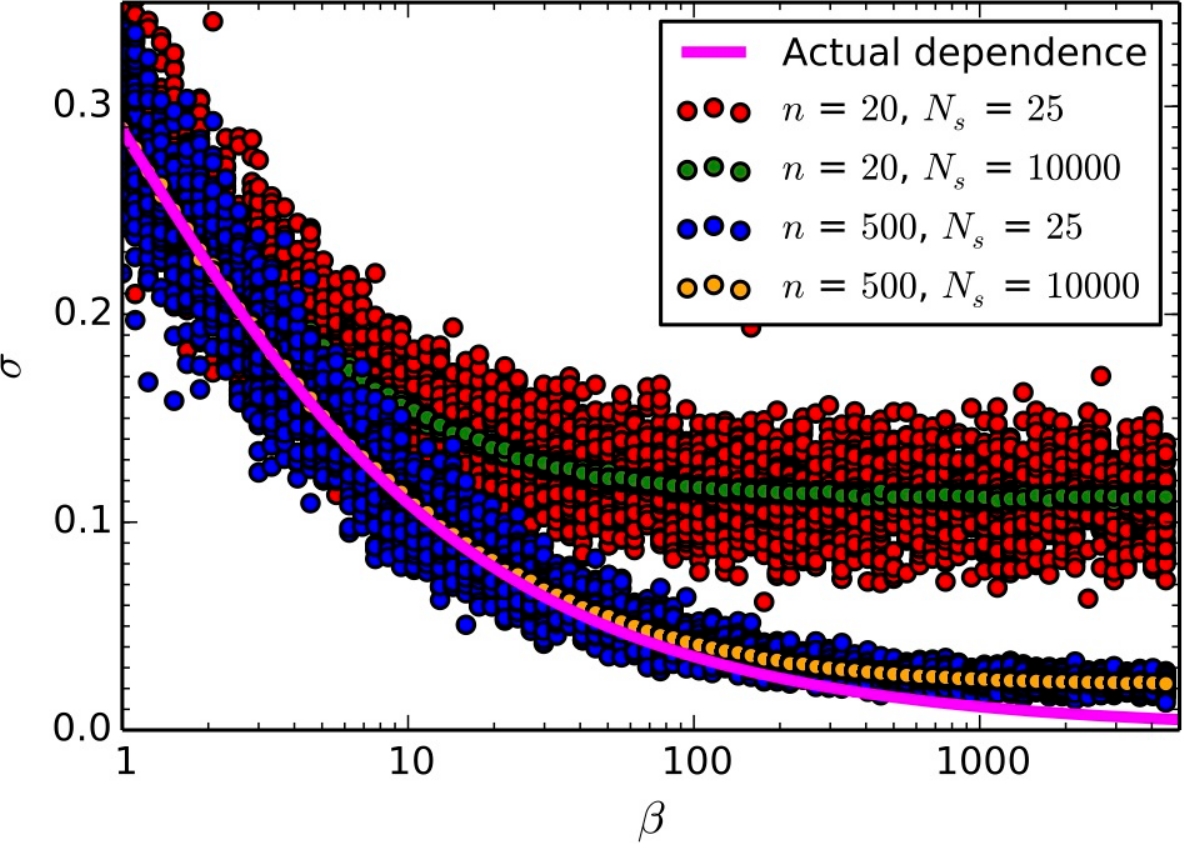}
   \caption{\textit{Symbols:} Randomly sampled distribution widths for 4 sets of parameters $N_s, n$ as a function of $\beta$. For each parameter combination, 50 random samples are plotted. \textit{Smooth curve:}  $\beta(\sigma)$ given by equation~(\ref{Eqn:BetaWidth}).
   \label{ModelingSamples}}
\end{figure}

For the following analysis, we will assume a shape for these PDFs.
It has been suggested that, at least in the atmospheric context~\citep{1995JSP....78..311S}, symmetric beta-distributions provide a good approximation for multiplier distributions of dissipation.
Such distributions have also been used in previous studies of particle concentrations~\citep[e.g.,][]{2007PhRvE..75e6305H} and are defined by
\begin{equation}
\label{Eqn:BetaPDF}
  f(m;\beta) = \left( m - m^2 \right)^{\beta-1} \frac{\Gamma(2\beta)}{2\Gamma(\beta)}
\end{equation}
with $\Gamma$ being the Gamma function.
The parameter $\beta$ determines the width of the distribution with small values of $\beta$ corresponding to wide, i.e. more intermittent, distributions. 
The width of the $\beta$ distribution (its standard deviation) is given by
\begin{equation}
\label{Eqn:BetaWidth}
  \sigma(\beta) = \sqrt{\frac{1}{4\left( 2\beta + 1 \right)}}. 
\end{equation}
However, similarly to the tracers, the \textit{observed} distribution width, $\sigma_0(\beta)$, will not only depend on the underlying $\beta$ value but also on the number of particles in a given sample, $n$, and the number of samples, $N_s$, used to compute the distribution.
In order to characterize this dependence, we have conducted Monte-Carlo experiments.
They mimic the finite-particle-number effects in the DNS under the assumption that the underlying probability distributions are $\beta$ distributions.
The procedure works in the following way:
First, for a given value of $\beta$, we draw a random multiplier $m$ from the $\beta$ distribution.
Given the number of particles $n$ in a sample volume (a given box), this corresponds to a partition into $n_\textrm{left} = mn$ and $n_\textrm{right} = (1-m)n$ particles for the two halves of the sample volume, where $n_\textrm{left}$ and $n_\textrm{right}$ are non-integers in general.
Then, we draw a random particle number $k_\textrm{left}$ from the corresponding Poisson distribution with expectation value $E(k_\textrm{left}) = n_\textrm{left}$.
This value (and $k_\textrm{right} = n - k_\textrm{left}$) represent one random sample of the number of particles found in two halves of a box, and correspond to ``observed'' multiplier values $m_\textrm{left}=k_\textrm{left}/n$ and $m_\textrm{right}=k_\textrm{right}/n$.
We repeat the procedure many times and compute the standard deviation $\sigma$ from all random samples $m_\textrm{left}$ and $m_\textrm{right}$ combined ($N_s$ total number of samples).
The result is a random sample of the ``observed'' distribution width given $n$, $N_s$, and the true underlying value of $\beta$.
Figure~\ref{ModelingSamples} shows the results from such experiments for selected parameters $n$, $N_s$. 
As one would expect, the scatter is large for a small number of samples, $N_s$, and becomes smaller with increasing $N_s$.
Also, one can see that a small number of particles per box, $n$, causes the observed distribution width to be systematically larger than the value from equation~(\ref{Eqn:BetaWidth}) (red and green symbols), but approaches the exact value as the number of particles gets large (blue and orange symbols).

Now that we understand how we can model the small-number statistics effects, we can proceed to correct the observations.
Given values of $n$ and $N_s$, we conduct the above described Monte-Carlo experiments for many random values of $\beta$, and find the value of $\beta$ that results in a distribution width closest to an observed width $\sigma_0$.
This gives us an estimate of the true underlying $\beta$ value.
In practice, we consider random values of $\beta$ between $1$ and $\infty$ by drawing samples with equal probability with respect to their true width $\sigma(\beta)$ (equation~\ref{Eqn:BetaWidth}).
Once we have accumulated at least $100$ samples resulting in distribution widths falling in a tolerance range between $\sigma_0 (1+\Delta)^{-1}$ and $\sigma_0 (1+\Delta)$, we compute the appropriate mean and standard deviation of those $\beta$ values.
Initially, a large value of $\Delta$ is chosen, but as more and more samples are accumulated, $\Delta$ is reduced step by step by factors of $2$ until the standard deviation converges, i.e. does not change significantly if $\Delta$ is reduced further.
We also require that the final $\Delta$ is no more than $0.01$ (the width is matched to within at least $1\%$ of the measured width).
To speed up the process, once $\Delta \le 0.1$, we restrict the range of $\beta$ samples drawn to values between $\bar{\beta} e^{ -2 \sqrt{\text{VAR}(\ln\bar\beta)}}$ and $\bar{\beta} e^{2 \sqrt{\text{VAR}(\ln\bar\beta)}}$ where $\bar\beta$ and $\sqrt{\text{VAR}(\ln\beta)}$ are the mean value and standard deviation of $\beta$ in the tolerance range.
Specifically, we define $\bar{\beta}$ as the $\beta$ value of the distribution whose variance (width squared) is the same as the arithmetic mean of the individual variances, i.e.  
\begin{equation}
  \label{Eqn:MeanBeta}
     \bar{\beta} = \beta(\bar{\sigma})
\end{equation}
with 
\begin{equation}
   \label{Eqn:MeanSigma}
   \bar{\sigma}^2 = \frac{1}{M} \sum_{i=1}^{M} \sigma(\beta_i)^2
\end{equation}
using equation~(\ref{Eqn:BetaWidth}) and its inverse $\beta(\sigma^2) = (8\sigma^2)^{-1}-\frac{1}{2}$, and where $\beta_i$ are the $\beta$ values of all the samples having widths within the tolerance range around $\sigma_0$.
The uncertainty in $\bar\beta$ is measured by its variance:
\begin{align}
    \text{VAR}(\ln\bar\beta) = \frac{\text{VAR}(\bar\beta)}{\bar\beta} \approx \nonumber \\
    \left( \frac{\partial \beta}{\partial \sigma^2}(\bar\sigma^2) \right)^2 \text{VAR}(\bar\sigma^2) = \frac{\text{VAR}(\bar\sigma^2)}{64 \bar\sigma^4}, 
\end{align}
following non-linear uncertainty propagation truncating the series after the first order, and where the variance of $\bar{\sigma}^2$ is computed by:
\begin{equation}
    \text{VAR}(\bar{\sigma}^2) = \frac{1}{M} \sum_{i=1}^{M} \left(\bar{\sigma}^2 - \sigma(\beta_i)^2 \right)^2.
\end{equation}
Once converged, the final values of $\bar\beta$ and $\text{VAR}(\ln\bar\beta)$ are estimates of the true $\beta$ value corrected for finite-particle-number effects, and an estimate of its uncertainty.

\subsubsection{Volume-averaged $\beta$-distributions}
\label{Subsubsection:MainResult}

\begin{figure*}
   \centering
   \vspace{0.0cm}
   \includegraphics[width=0.9\linewidth]{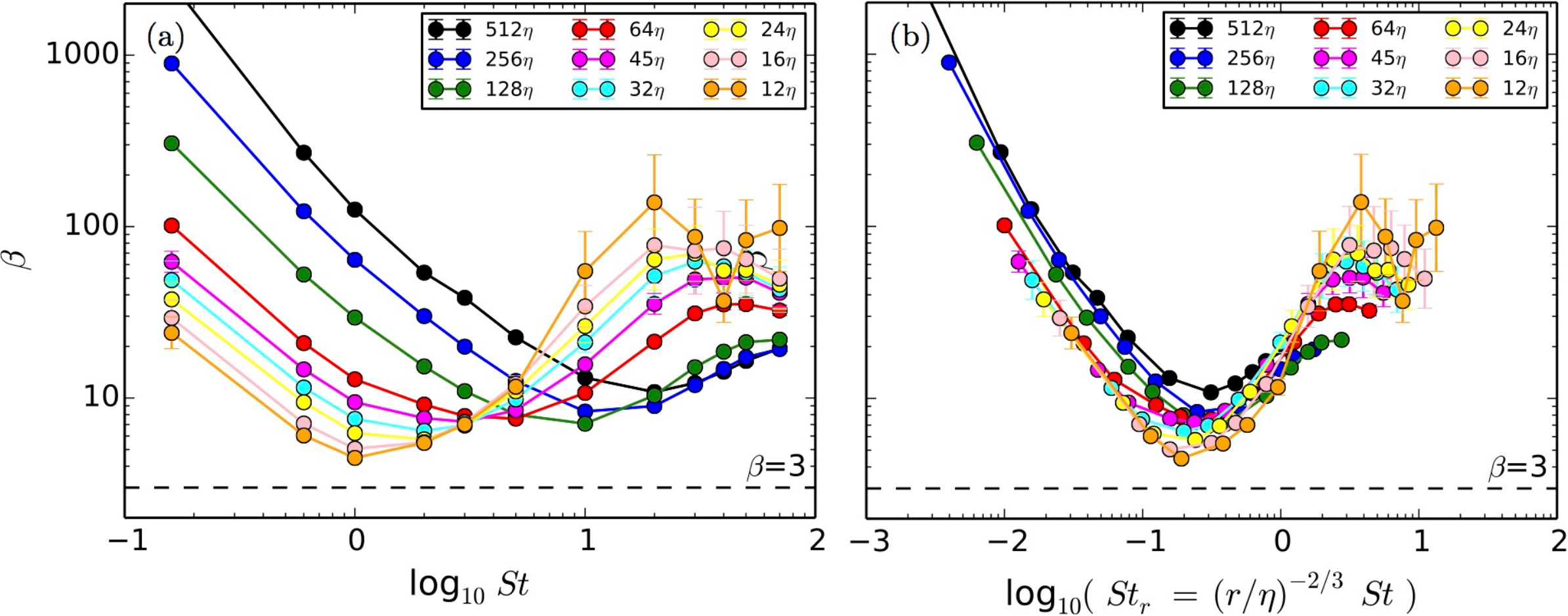}
   \caption{Concentration multiplier distribution $\beta$ values corrected for finite-particle number sampling effect as a function of the ordinary Stokes number $St$ (a), and in terms of the \textit{local} Stokes number $St_r$ (b). Solid lines connect results for the same binning scale for guiding the eye.  
   \label{Fig:MultiplierBetas}}
\end{figure*}

Now that we know how to remove the effects that a finite number of particles has on the multiplier distributions for a given box size or scale, we compute corrected volume averages of the $\beta$ values for any given spatial scale and Stokes number.
Specifically, at any given scale $r$, we first compute the multiplier values for all the sampling boxes.
We then compute the distribution widths $\sigma_{0,j}$ from each subset of boxes having the same number of particles $n_j$.
This width is then used to derive a corrected $\beta$ value, which corresponds to a corrected width $\sigma_j$ (through equation~\ref{Eqn:BetaWidth}).
We then combine all results by summing the square of these corrected $\sigma_j$ values weighted with the fractional volume that the boxes with $n_j$ particles occupy.
The final, combined $\beta$ value is then given by this average width through the inverse of~(\ref{Eqn:BetaWidth}).

Figure~\ref{Fig:MultiplierBetas}a shows the combined concentration multiplier $\beta$ values for all Stokes numbers and all spatial scales considered.
We did not consider scales smaller than $12\eta$ since the number of particles in such small boxes is too small to make reliable statistical inferences, even with our correction procedure.
From the results, it is very apparent that the concentration multiplier distributions are not only functions of $St$, as has been known, but are also very much scale dependent.
It seems intuitive, however, that at any given scale, the multipliers may depend only on the ratio of the particle stopping time and the dynamical time at that spatial scale.
An effect of this sort was seen in particle concentration PDFs by~\citet{2007PhRvL..98h4502B}.
Along these lines, we construct a \textit{local} Stokes number
\begin{equation}
   St_r \equiv \frac{\tau_s}{\tau_r} = St \left(\frac{r}{\eta}\right)^{-2/3}, 
   \label{Eqn:LocalStokesNumber}
\end{equation}
where we have assumed that the dynamical time at scale $r$ is given by
\begin{equation}
 \tau_r = \tau_\eta \left(\frac{r}{\eta}\right)^{2/3}
 \label{Eqn:TauR}
\end{equation}
following~\citet{1962JFM....13...82K}.
By plotting the multiplier $\beta$ results against the rescaled Stokes number, the curves approximately collapse into one (figure~\ref{Fig:MultiplierBetas}b), at least for scales not too close to the integral scale.
This means that when local scale and stopping time are accounted for, the \textit{scaled} multiplier $\beta$ curves are cascade level \textit{independent} for scales $r < L/10$ or so, and are thus highly amenable to cascade models at, in principle, arbitrarily large $Re$ at least within the inertial range.
For some caveats about $Re$-dependence however, see section~\ref{Section:Discussion}.

Also note that $St_r$ can be written in terms of an \textit{integral-scale} Stokes number $St_L$:
 \begin{equation}
 St_r = 2^{2N/9} St_L,
  \label{Eqn:Local_Stokes_Rewritten}
 \end{equation}
 where
 \begin{equation} 
 St_L \equiv \frac{\tau_s}{\tau_L} = St \left(\frac{L}{\eta}\right)^{-2/3}
 \label{Eqn:IntegralScaleStokesNumber}
 \end{equation}
using the same Kolmogorov scaling as in equation~(\ref{Eqn:TauR}).
Equation~(\ref{Eqn:Local_Stokes_Rewritten}) separates terms that depend only on cascade level (\textit{first term}), and particle properties (\textit{second term}), and indicates that particles in different flows behave the same (have the same statistical cascade) if they have the same integral-scale Stokes number $St_L$.
We will use this fact later in section~\ref{Subsection:ComparisonCascadeWithDNS} when we compare the cascade with DNS results from two different simulations at different $Re$.

It should be noted our observed $r^{-2/3}$-scaling seems to contradict the scaling found by~\citet{2007PhRvL..98h4502B} for ``quasi-Lagrangian'' probability distribution functions of the mass density.
Following an idea by~\citet{Maxey1987}, they approximated the dynamics of inertial particle by those of tracers in an appropriate synthetic compressible velocity field and derived a scaling for the rate at which an $r$-sized ``blob'' of particles contracts.
They argued that the scaling of the contraction rate relates to the scaling of the pressure field, give the contraction rate as being proportional to $r^{-5/3}$ for the $Re$ of their simulation~\footnote{They argue that the scaling should change to $r^{-4/3}$ for very high Reynolds numbers $Re_\lambda \ge 600$}, and find that their density PDFs collapse when scaled with this contraction rate.

\subsubsection{Composite PDFs}
\label{Subsubsection:CompositePDFs}

\begin{figure*}
   \centering
   \includegraphics[width=0.9\linewidth]{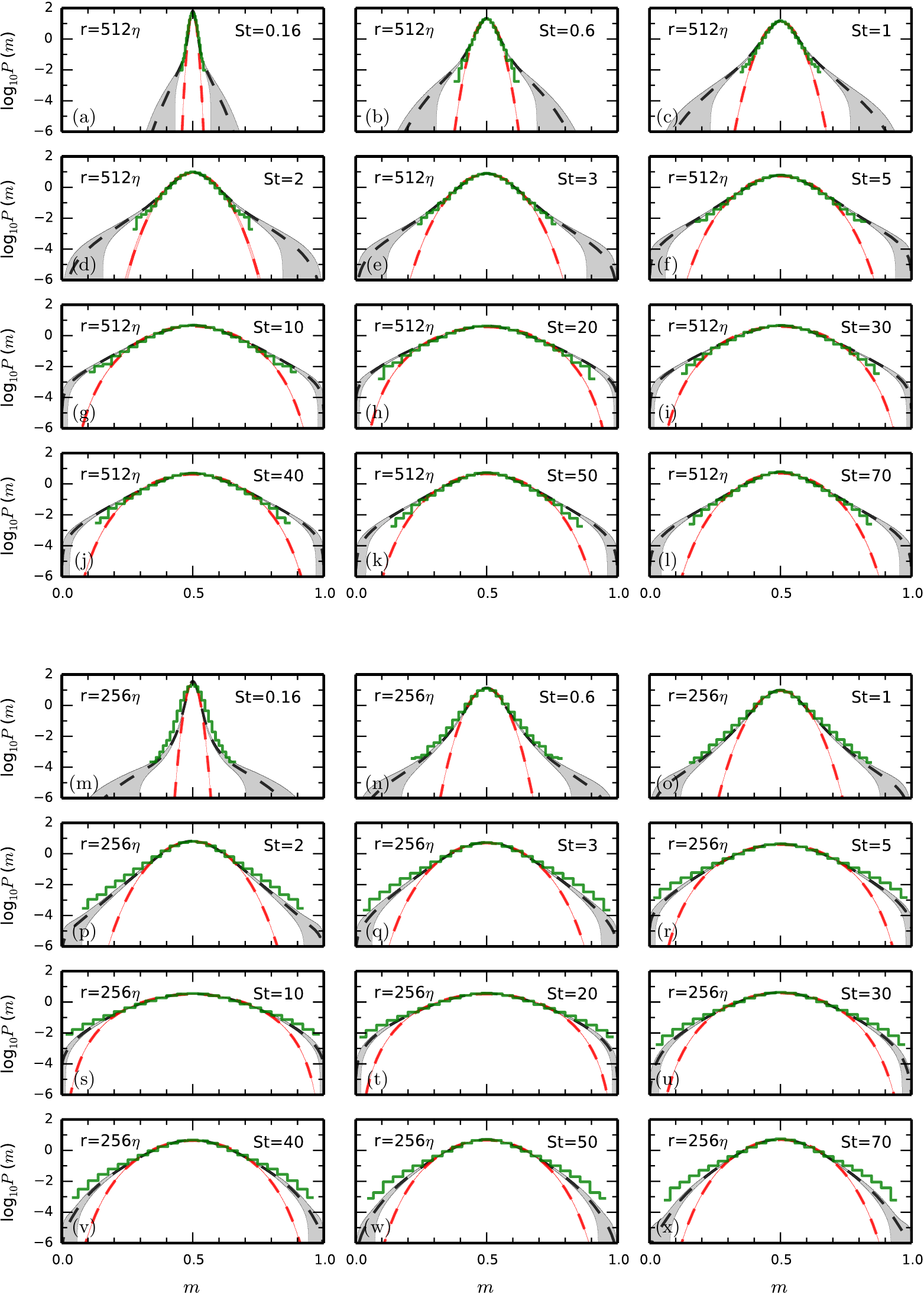}
   \caption{\label{PDFs512and256} Probability distribution functions (PDFs) of particle multipliers measured from DNS particle trajectories for different Stokes numbers and a box size of $r=512\eta$ (a through l) and $256\eta$ (m through x). Shown are the volume-averaged, mean $\beta$-distributions (\textit{red dashed curves\/}) as described in section~\ref{Subsubsection:MainResult}, the composite PDFs (\textit{black dashed curves\/}) as defined in section~\ref{Subsubsection:CompositePDFs}, and raw multiplier histograms uncorrected for finite-particle-number effects (\textit{green curves\/}). Corresponding shading in red and grey shows the uncertainty in the measured PDFs (plus/minus one standard deviation).}
\end{figure*}

\begin{figure*}
   \centering
   \includegraphics[width=0.9\linewidth]{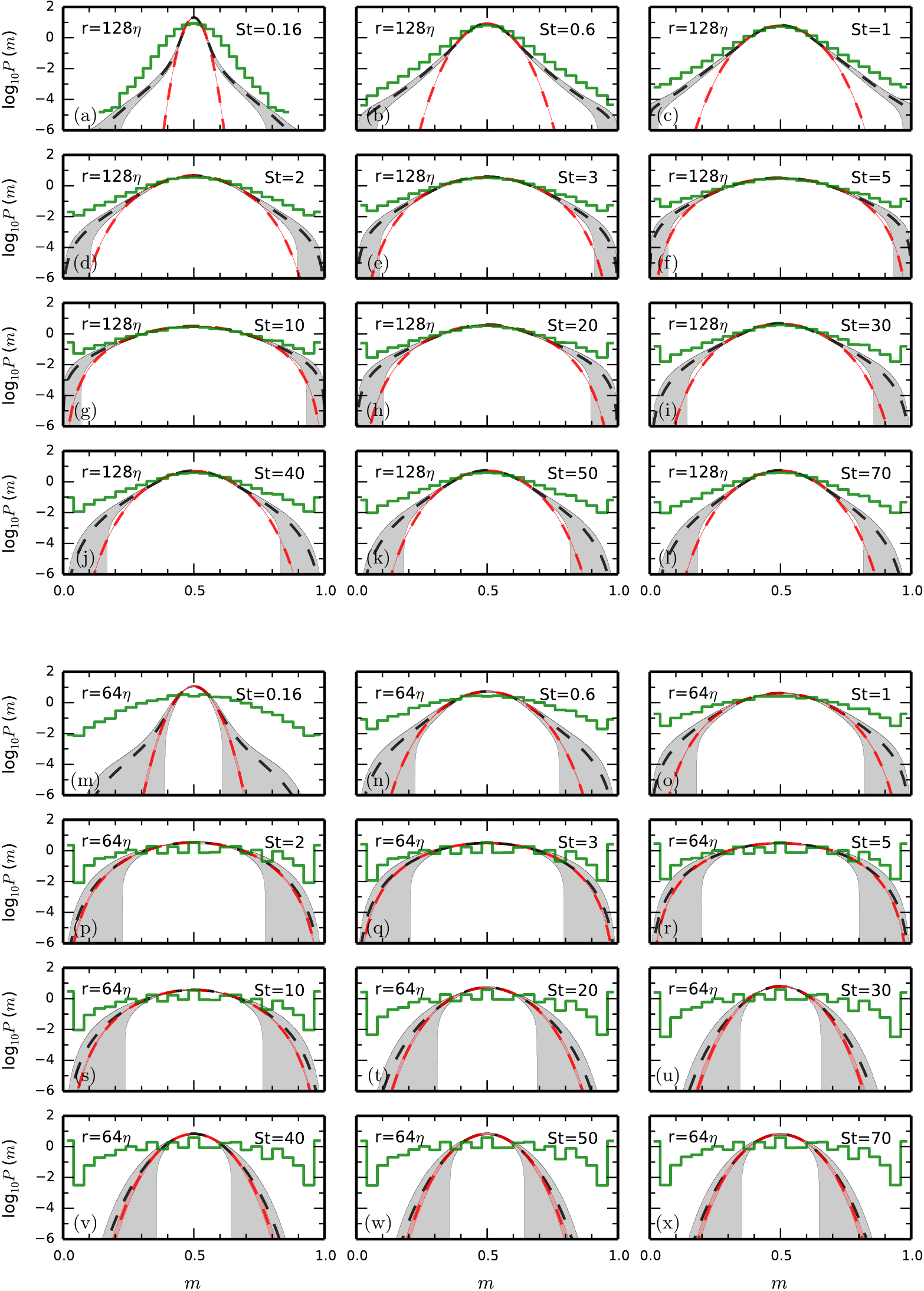}
   \caption{\label{PDFs128and64}Same as figure~\ref{PDFs512and256} but for box sizes $r=128\eta$ (a through l) and $64\eta$ (m through x).}
\end{figure*}

\begin{figure*}
   \centering
   \includegraphics[width=0.9\linewidth]{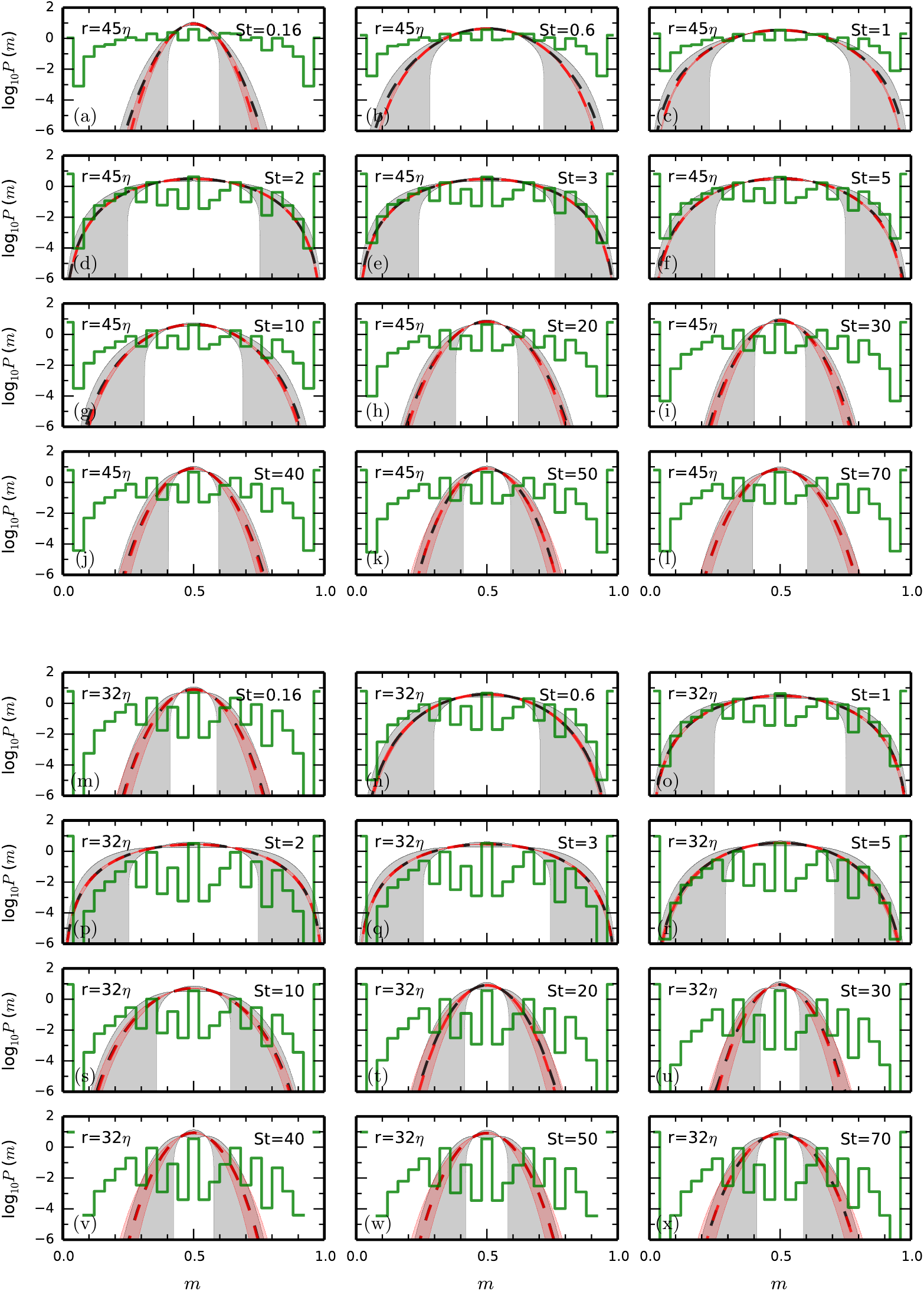}
   \caption{\label{PDFs45and32}Same as figure~\ref{PDFs512and256} but for box sizes $r=45\eta$ (a through l) and $32\eta$ (m through x).}
\end{figure*}

Combining the statistics from boxes with different particle number to form a single $\beta$-distribution that represents the concentration multipliers assumes that multiplier distributions are the same for all concentrations.
Since the particles are independent, that is, they do not feel the presence of each other, one would assume that there is no such concentration dependence.
However, particles concentrate in particular regions of the flow, and therefore flow properties differ in regions of different particle density and so the multiplier distributions may also be different.
We can relax the assumption of equal multiplier distributions and compute a \textit{composite} multiplier distribution by first computing $\beta$-distributions for each particle concentration separately, and then summing these distributions weighted with the fractional volume that the boxes with the particular density occupy.
These composite distributions are shown in figures~\ref{PDFs512and256}, \ref{PDFs128and64} and~\ref{PDFs45and32}.

The following observations can be made:
First, composite and mean $\beta$-distributions may differ in shape, although by construction they have identical widths (second moments).
That is, in general, the composite $\beta$-distribution is itself \textit{not} a $\beta$-distribution.
At large spatial scales, \textit{flat} or exponential tails are apparent in the composite PDFs; these do not have the same shape as \textit{any} $\beta$-distribution.
At small scales, however, differences disappear within the margin of accuracy.
Also, at the largest scale ($512\eta$) there are enough particles such that finite-particle-number effects are small and the raw PDFs (shown in green) are essentially identical to the composite PDFs (shown in black).
However, the importance of correcting for sampling effects becomes apparent as we look at smaller scales where fewer particles cause spurious widening of the raw distributions or ``false intermittency.''
Finally, we should mention that at the smallest scales, the number of particles is so small that we are only measuring multipliers in high-concentration regions, which causes a sampling bias since particles are known to avoid vorticity, and such regions may produce more intermittency or broader multiplier PDFs.
For instance, the average number of particles in a $32\eta$ box, for the Stokes numbers for which we have a total of $6\times10^5$ particles (see section~\ref{Section:Dataset}) is only $\left( 32\eta/2048\eta \right)^3 N_p \approx  0.44$.
The situation is even worse for, say, $r=12\eta$ and a case with only $2\times10^5$ total particles.
The average is then only 0.04 particles per sampling box.
Multipliers are measured only in regions with a particle concentration that is at least 100 times larger than the mean concentrations since we can only reasonably measure multipliers if we have at least several particles in a sampling box.

\subsection{Dissipation and enstrophy multiplier distributions}
\label{Subsection:Analysis:DissipationAndEnstrophy}

\begin{figure}
   \centering
   \includegraphics[width=1.0\linewidth]{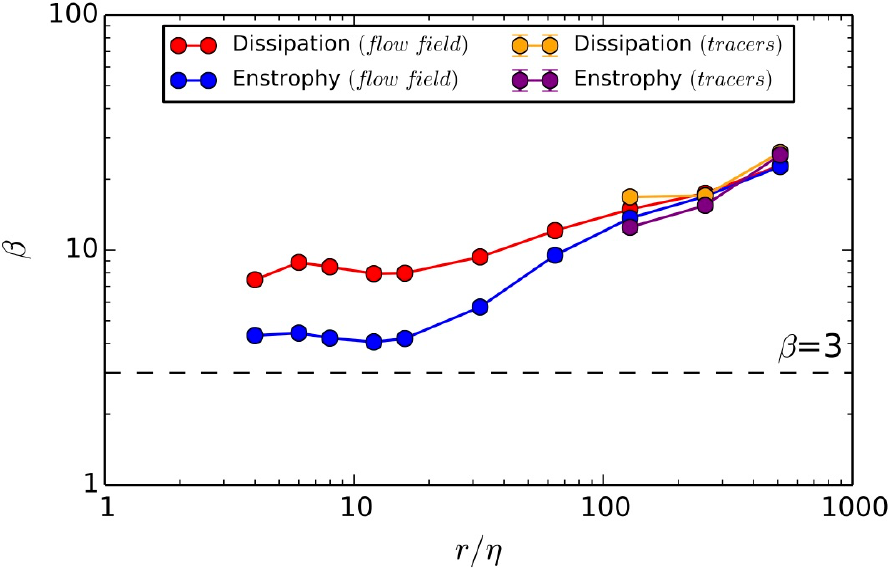}
   \caption{
Dissipation and enstrophy multiplier distribution $\beta$-values as a function of spatial scale computed from the flow data and from tracer particle trajectories ($St=0$). Results from tracer particles are only shown for large spatial scales since at smaller scales not enough particles are available for a reasonable estimation of the multiplier distributions.
   \label{Fig:DissipationEnstrophyBeta}}
   \end{figure}
   
We also calculated the multipliers for fluid dissipation and enstrophy.
The rate of turbulent dissipation is given by
\begin{equation}
   \epsilon = 2 \nu S_{ij} S_{ij} = \nu \left[ (\partial_i u_j) (\partial_i u_j) + (\partial_i u_j) (\partial_j u_i)  \right],
\end{equation}
where $S_{ij}$ and $u_i$ are the strain rate tensor and the components of the velocity field, respectively, and where we use the Einstein summation convention.
The enstrophy on the other hand is defined as the square of the vorticity:
\begin{equation}
   {\cal E} = |\nabla \times \vec{u} |^2 = (\partial_i u_j) (\partial_i u_j) - (\partial_i u_j) (\partial_j u_i).
\end{equation}
All of these flow velocity derivatives are available in the dataset, both in the particle trajectory data files and in the flow snapshots.
We have computed multiplier distribution $\beta$ values for $\epsilon$ and $\cal E$ ($\beta_\epsilon$ and $\beta_{\cal E}$) from the trajectory of tracer particles (as they sample the flow more homogeneously than non-zero Stokes number particles), and from the full resolution flow snapshots.
The results are shown in figure~\ref{Fig:DissipationEnstrophyBeta}, although the tracer data is only shown for the largest spatial scales since it also suffers from finite-particle effects (not corrected here).

Note the presence of asymptotes for $r \lesssim 20\eta$ (perhaps better thought of as $r \lesssim L/50$, see figure~\ref{fig:hypothesis}) for both, as anticipated~\citep{1995JSP....78..311S}.
Enstrophy is shown to have wider multiplier distributions (smaller $\beta$ values) than dissipation, and is therefore more intermittent.
This is consistent with the findings of~\citet{1990PhRvA..41..894M} in several flows including atmospheric flow, and in numerical simulations~\citep[e.g.,][]{1996PhRvL..77.3799C, 1997PhRvL..79.1253C}. For a review, see~\citet{1997AnRFM..29..435S}.
Also, we note that even for the smallest spatial scales considered, still well within the inertial range, the dissipation rate multiplier $\beta$ does not reach the atmospheric flow values of $\beta_\epsilon \sim 3$~\citep{1995JSP....78..311S}.
See section~\ref{Section:Discussion} for more discussion.
 
\section{New cascade model with level-dependent multipliers}
\label{Section:NewCascade}

\subsection{Cascade simulations}
\label{Subsection:CascadeSimulation}

\begin{figure}
   \centering
   \includegraphics[width=0.9\linewidth]{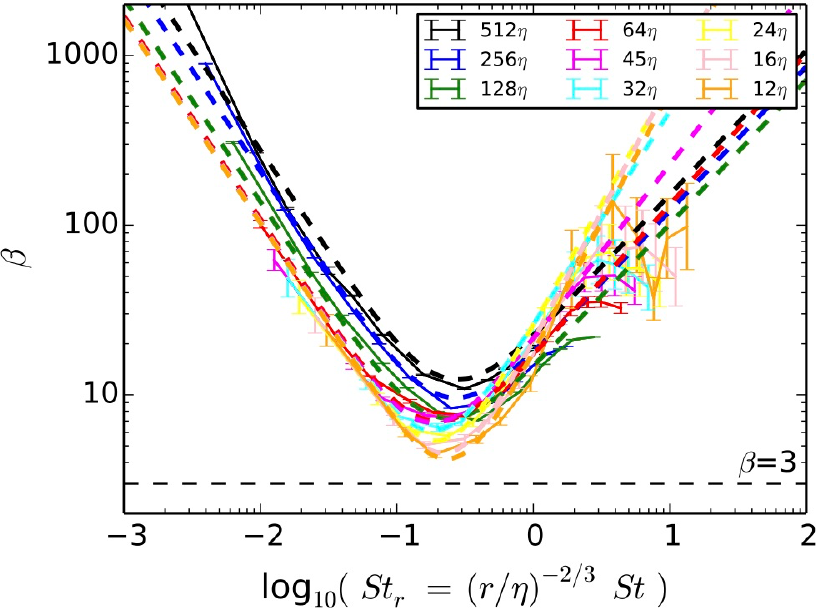}
   \caption{Beta values of the cascade model following equations~(\ref{Eqn:BetaModel}) and~(\ref{Eqn:Parameters}) (\textit{dashed lines\/}). Results from the present DNS data are shown for comparison (\textit{solid lines}), they are the same curves as in figure~\ref{Fig:MultiplierBetas} except that the symbols have been suppressed here for legibility. Spatial scales, $r$, are differentiated by color (see figure legend).
   \label{Fig:NominalModel}}
\end{figure}

From the collapsed $\beta(St_r)$ curves (figure~\ref{Fig:MultiplierBetas}b), we can build an empirical model for the particle multiplier distributions.
A sum of two power laws approximates the curves for fixed scale $r$ well:
\begin{equation}
   \beta(St_r) \approx \beta_\textrm{min} \left(  \left( \frac{St_r}{a_1} \right)^{b_1}  +   \left( \frac{St_r}{a_2} \right)^{b_2}   \right),
   \label{Eqn:BetaModel}
\end{equation}
with parameters $a_1$, $a_2$, $b_1$, $b_2$ determining the slopes and positions of the exponentials, respectively, and $\beta_\textrm{min}$ setting the minimum $\beta$ value.
From the figure it is evident that there is some residual scale dependence -- the curves for different spatial scales don't overlap exactly.
The following parameterization approximates this residual dependence: 
\begin{align}
   & a_1 = 0.15, & \nonumber \\
   & a_2 = 0.45 - 0.25 \exp\left( - \frac{2}{3} \ln \left( 20.5~\frac{r}{L} \right)^2 \right),  & \nonumber \\
   & b_1 = -1.2, & \nonumber \\
   & b_2 = 0.85 + 0.35 \left[1+\erf\left( - 1.8 \ln\left( 29.3~\frac{r}{L} \right) \right) \right], &  \nonumber \\
   & \beta_\textrm{min} = 4 + 4 \left[ 1+\erf\left( \ln\left( 4~\frac{r}{L} \right) \right) \right], &
   \label{Eqn:Parameters}
\end{align}
where $\erf$ and $\ln$ are the error function and the natural logarithm.
The parameters asymptote for small $r/L \lesssim 3\times10^{-3}$ (or large cascade level $N \gtrsim 25$) to:
\begin{align}
   & a_1 = 0.15, \,\,\, a_2 = 0.45, \nonumber \\
   & b_1 = -1.2, \,\,\, b_2 = 1.55, \,\,\, \beta_\textrm{min} = 4. &
   \label{Eqn:AsymptoticParameters}
\end{align}
The model is shown in figure~\ref{Fig:NominalModel} compared to the DNS results.
An even simpler model could probably be constructed using a single average curve of all $\frac{r}{L} \le \frac{1}{8}$.

With this model for the particle multiplier distributions, we can perform statistical cascade simulations to predict the probability distribution function for the particle concentration.
We start at cascade level 0 with a single concentration value of $C^{(0)}=1.0$.
At every cascade level $N$, we then draw a random multiplier value $m^{(N)}$ from the corresponding distribution with a $\beta$ value given by the model (equation~\ref{Eqn:Parameters}), and split the concentration value from the previous level into two values $C^{(N)}_\textrm{left} = 2 C^{(N-1)} m^{(N)}$ and $C^{(N)}_\textrm{right} = 2 C^{(N-1)} (1-m^{(N)})$.
The factor 2 here comes from the fact that the concentration is the ratio of the particle number in a half box and the mean number in a half box (which itself is one half of the mean number of particles in full box).
Such a cascade produces $2^N$ random samples of concentrations values at each cascade level $N$ which we use to compute concentration PDFs. For good statistics, however, we need many more samples, and for the predictions shown in the following section we computed 50,000 such cascade simulations.

\subsection{Level-dependent cascade predictions compared to DNS at two different $Re$}
\label{Subsection:ComparisonCascadeWithDNS}

\begin{figure*}
   \centering
   \includegraphics[width=0.85\linewidth]{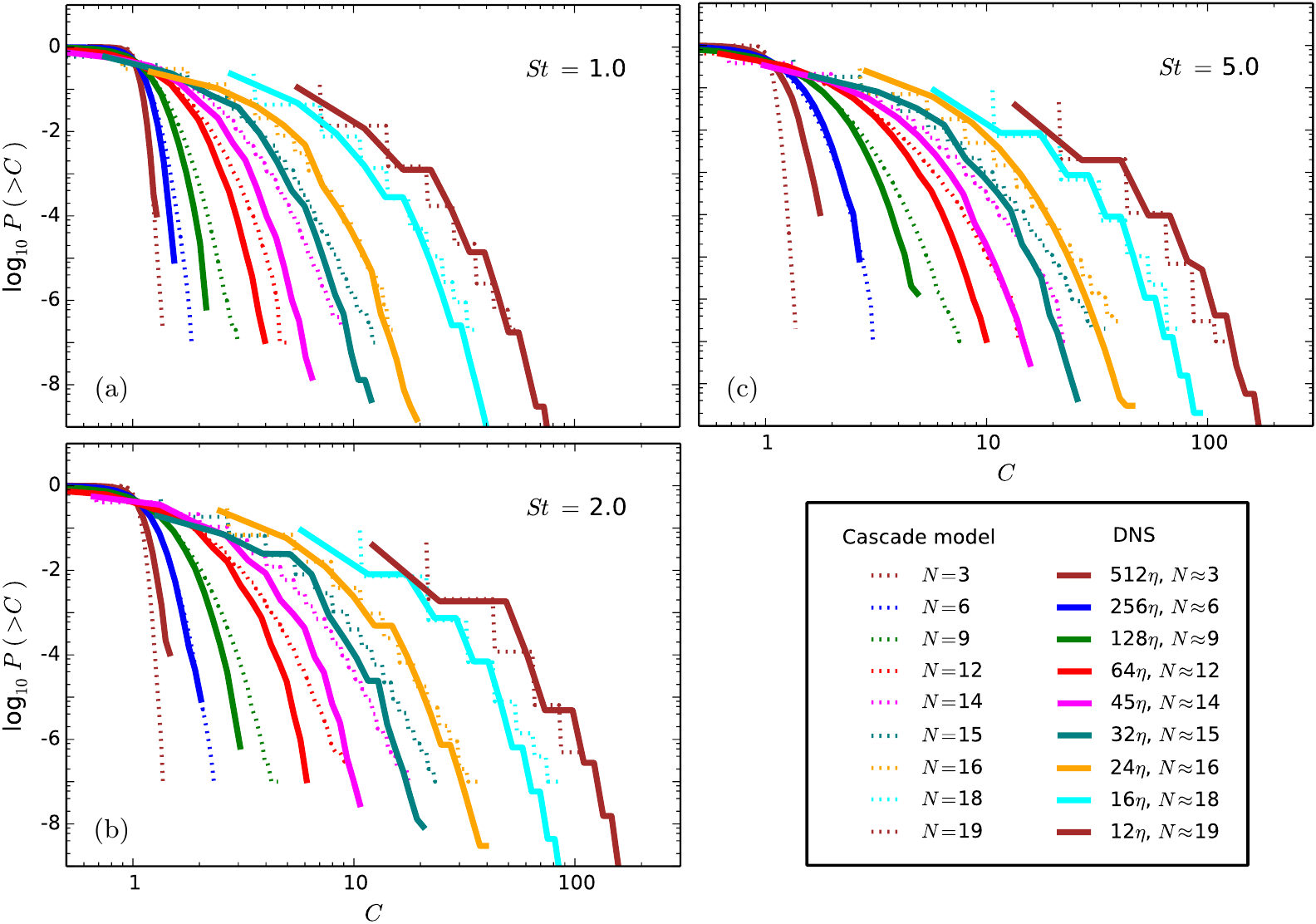}
   \caption{\label{Fig:ComparisonCascadeWithOurDNS} 
   Cumulative probability distribution functions for the concentration factor $C$, for different Stokes numbers, comparing the DNS measurements (\textit{solid curves\/}) with the cascade predictions, degraded to account for finite-particle-number effects (\textit{dotted curves\/}).}
\end{figure*}

In order to demonstrate and assess the cascade predictions, we compare the probability distribution functions of the concentration factor generated by the cascade model with those measured directly from DNS datasets. 
In order to do so, we need to account for the small-number effects present in the DNS results which, as we have seen before, can cause observed distributions to be significantly widened relative to \textit{ideal} ones that the cascade produces.
Instead of \textit{correcting} the DNS PDFs as we have done before for the measured multiplier distributions, we will \textit{degrade} the cascade PDFs for this purpose by introducing finite-particle-number effects into them.

For motivating the procedure, let us imagine a hypothetical simulation $\cal H$ with a number of particles so large that finite-particle-number effects are negligible, and let $\bar{n}_\infty(r)$ be the average number of particles in a sampling box at length scale $r$ in that simulation. 
Our present dataset, in comparison, has on average only $\bar{n}(r) = N_p r^3 / {\cal L}^3$ particles in a box of scale $r$, where as before $N_p$ is the total number of particles in the dataset with a given Stokes number, and $\cal{L}$ is the linear extent of the simulation domain, respectively.
One can think of our current dataset as a randomly selected subset of the hypothetical simulation $\cal H$, generated by retaining particles from $\cal H$ with a probability of $p(r)=\bar{n}(r)/\bar{n}_\infty(r)$.
For brevity, we will suppress the $r$ below. 
Specifically, let's say some sampling box in $\cal H$ has $n_\infty$ particles in it (and therefore a concentration factor $C_\infty$ = $n_\infty / \bar{n}_\infty$).
From these, we select particles with a probability of $p$, retaining in total $n$ particles, where $n$ is an integer random number with an expectation value of $E(n) = C_\infty \bar{n}$. 
For $\bar{n}_\infty \rightarrow \infty$, this is a Poisson process and $n$ is a random number with a probability mass function
\begin{equation}
P_P(n;C_\infty \bar{n}) = \frac{(C_\infty \bar{n})^n e^{-C_\infty \bar{n}}}{n!}.
\end{equation}

Using this idea, the recipe for introducing finite-particle-number effects into the cascade PDFs is as follows:
First, we draw a random sample $C_\infty$ from a cascade-derived concentration PDF.
Second, we draw a random sample $n$ from a Poisson distribution with the corresponding expectation value $E(n) = C_\infty \bar{n}$, where $\bar{n}$ is again the average number of particles at the spatial scale of interest in the DNS data we want to compare. 
The (integer) particle number $n$ corresponds to a discrete concentration factor $C = n / \bar{n}$.
By repeating the procedure $N_s$ times, we can build from the samples a discrete probability distribution function of $C$.
It accounts for the finite-particle-number effects and can be directly compared to PDFs measured from the DNS dataset.

Figure~\ref{Fig:ComparisonCascadeWithOurDNS} shows, for different Stokes numbers, a comparison between the PDFs predicted by the cascade model, and degraded in this way using $N_s = 10^7$, with those calculated directly from the DNS dataset we analyzed in this paper~\citep{RM-2007-GRAD-2048}.


\begin{figure}
   \centering
   \includegraphics[width=0.95\linewidth]{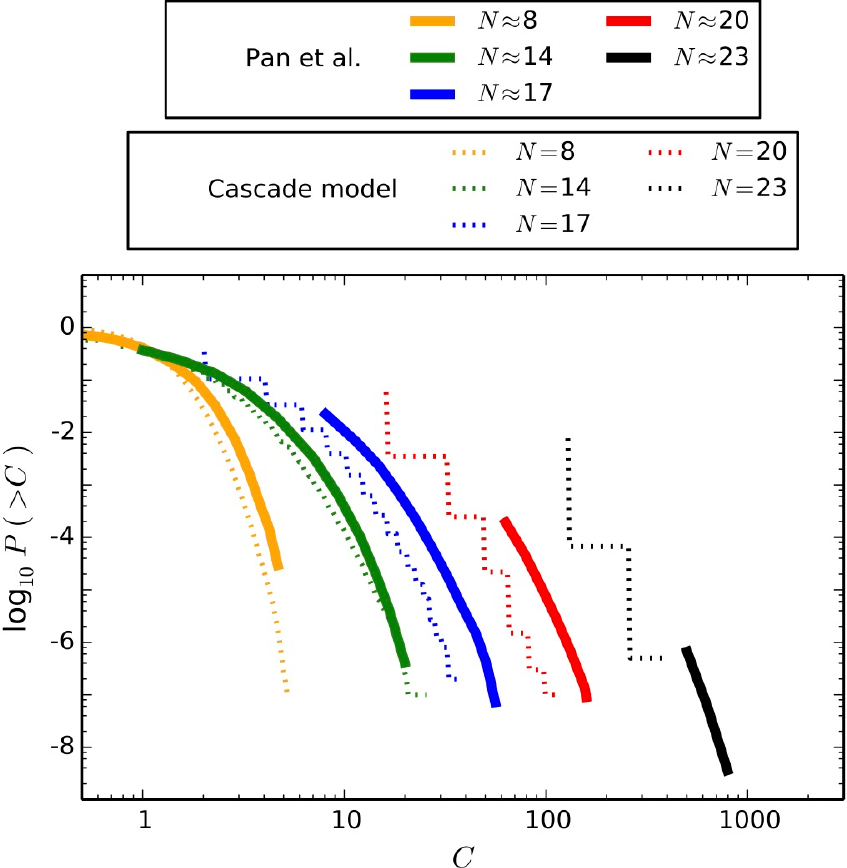}
   \caption{\label{Fig:ComparisonCascadeWithPan} 
   Cumulative probability distribution functions for the concentration factor $C$, for different cascade levels comparing DNS results from figure~8 of~\citet{0004-637X-740-1-6} with an estimated Stokes number $St \approx 1.2$ or $St_L \approx 3.5 \times 10^{-2}$ (\textit{solid curves\/}) with the prediction of our cascade model degraded to account for the finite-particle-number effects in Pan's DNS dataset (\textit{dotted curves\/}). For our cascade, we have used our $St = 5$ model with a $St_L \approx 5 \times 10^{-2}$  close to the value of Pan.}
\end{figure}

Under the assumption that the collapsed multiplier $\beta$ curve is universal and does not depend on Reynolds number, we can use cascade simulation to model conditions at different Reynolds numbers.
For caveats to this and a suggestion regarding plausible $Re$ dependence, see the discussion in section~\ref{Section:Discussion}.
It follows from equation~(\ref{Eqn:Local_Stokes_Rewritten}) that particles of different $Re$ flows behave the same and have the same cascade statistics, if they have identical integral-scale Stokes numbers $St_L$.

Here, we compare our cascade with~\citet{0004-637X-740-1-6} who performed direct numerical simulation of a compressible flow with suspended initial particles. 
Their simulation is on a $512^3$ Cartesian grid with an estimated $L/\eta \sim 200$ (compared to $L/\eta \sim 1024$ in the dataset we used here), and contains $8.6\times 10^6$ particles per Stokes number.
Initial comparison of their uncorrected multiplier PDFs with ours showed a clear disagreement but most of that disagreement disappears once we take into account finite-particle number effects.
Figure~\ref{Fig:ComparisonCascadeWithPan} shows a comparison of cumulative PDFs of the concentration between~\citet{0004-637X-740-1-6} and our cascade model showing reasonable agreement, bearing in mind that the inviscid simulations of~\citet{0004-637X-740-1-6} leave a little uncertainty about the value of $St$.

 \subsection{Model prediction for the radial distribution function}
 \label{Subsection:RDF}
 
In many applications, e.g. in terrestrial clouds, particle collisions play an important role, and it is therefore of great interest to model this process.
The rate of collisions depends on two statistical quantities: the radial relative velocity between particles, and the radial distribution function (RDF), $g(r)$, defined as the probability of finding two particles at a given separation normalized with respect to homogeneously distributed particles~\citep{0004-637X-740-1-6,BraggCollins2014I}.
Relative velocities are beyond the scope of the present model, but the cascade model can be used to make predictions for the RDF.
For this purpose, we performed statistical simulations similar to section~\ref{Subsection:CascadeSimulation} but with an important difference: we are here interested in the spatial distribution of the concentration, which then can be used to compute the RDF.

Our cascade model, however, only describes the particle multiplier PDF as a function of cascade level, and does not explicitly contain information about \textit{spatial correlations}.
There is therefore some ambiguity in how to compute concentration fields from the cascade model.
We have explored three different methods to assess the range of possible solutions.
Starting at the largest scale, we divide a cube of space in half along each spatial direction.
This results in eight sub-cubes with half the linear size.
In order to solve for the concentrations in these sub-cubes uniquely, we need eight equations. 
The first method -- \textit{method A} --  makes the following choice: one constraint it given by the fact that the average of the concentrations over all eight sub-cubes is equal to the mean concentration, and seven additional constraints are given by relating the concentration in seven sets of neighboring sub-cubes through multipliers chosen randomly from the cascade model.
The specific choice of equations is given in appendix~\ref{Appendix}.
Solving for concentrations to ever smaller cubes until some small cutoff length scale $(r/L)_\textrm{min}$ then yields a statistical realization of the concentration field that can be used to compute the RDF.
The probability of finding two particles at a given distance is simply the product of the concentrations at the two points in space a given distance apart, averaged over the whole domain.
If we start with a unit concentration at the largest scale, the normalization, that is the probability for homogeneously distributed particles, is simply 1.
In order to reduce the computational cost and storage requirements to trackable amounts, we do not follow all sub-cubes to ever smaller scale but only a random selection of them. One half of sub-cubes are followed at each cascade level.
Two more methods are obtained by relating the concentrations in the two half-cubes, for each direction separately, through a random multiplier.
This set of equations is underdetermined.
For \textit{method B} we pick one particular solution, while for \textit{method C} we use a least-square solver to determine the minimum-norm solution.
For specifics, again, we refer to appendix~\ref{Appendix}.
Conceptually, it is clear that the three methods allow for different amounts of spatial randomness.
\textit{Method A} clearly maximizes intermittency while \textit{method C} leads to the least spatially intermittent solution.

\begin{figure*}
   \centering   
   \includegraphics[width=0.9\linewidth]{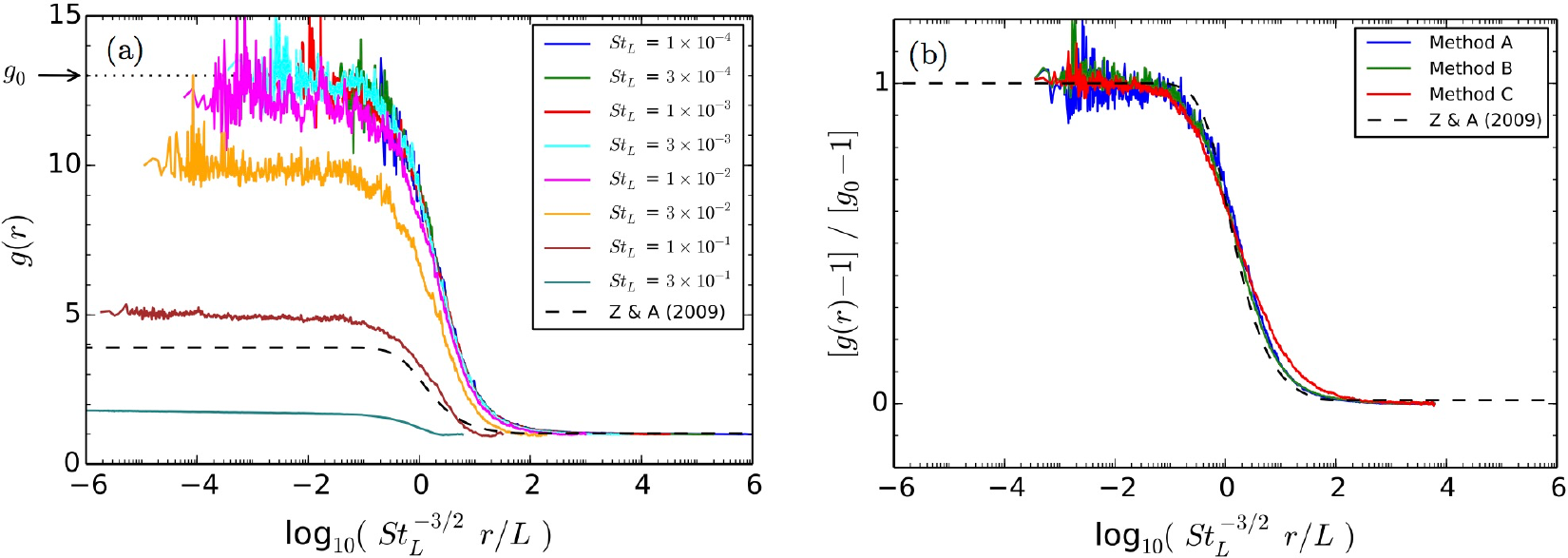}
   \caption{(a) Radial distribution function (RDF) $g(r)$ for various values of $St_L$ (\textit{coloured solid lines}) determined from cascade model simulations using method $A$ with the asymptotic RDF value, $g_0$, indicated by an arrow. For comparison, the model of~\citet{ZaichikAlipchenkov2009} from their figure~3 for infinite Reynolds number (\textit{black dashed line}) is also shown. (b) RDFs for $St_L = 3 \times 10^{-3}$ for the different cascade simulation methods, and the ~\citet{ZaichikAlipchenkov2009} model, scaled by their respective asymptotic value, $g_0$, given in table~\ref{Tab:AsymptoticValues}.
   \label{Fig:CascadeRDFs}}
\end{figure*}

Figure~\ref{Fig:CascadeRDFs} shows RDFs predicted by our cascade model simulations down to cascade level 72 (spatial scales of $(r/L)_\textrm{min} \approx 6 \times 10^{-8}$).
All methods give qualitatively the same results, all in good agreement with the~\citet{ZaichikAlipchenkov2009} asymptotic $Re=\infty$ analytical solution, especially regarding the shape and the active range of scales.
In fact, the magnitudes are even close enough, within a factor of order unity. 
RDFs for different $St_L$ computed using method A are plotted in figure~\ref{Fig:CascadeRDFs}a as a function of the scaled distance $St_L^{-3/2} r/L$, and are shown to collapse for small $St_L \lesssim 10^{-2}$.
Effectively, the scaling behavior of the multiplier PDFs (figure~\ref{Fig:MultiplierBetas}) is carried over to the RDF.
At large scales, $St_L^{-3/2} r/L \gtrsim 10^2$, the RDF has a value of 1, that is, particles are homogeneously distributed, and over an \textit{active} range of scales, $10^{-2} \gtrsim St_L^{-3/2} r/L  \gtrsim 10^2$, it rises and reaches an asymptote, $g_0$.
For larger $St_L$ in our sample, however, the active range is shortened by being too close to the integral scale $L$, and $g(r)$ asymptotes at smaller values that are $St_L$-dependent.
The theory of~\citet{ZaichikAlipchenkov2009} predicts a very similar behavior and their curve for infinite Reynolds number, effectively for infinitely small $St_L$, is also shown in figure~\ref{Fig:CascadeRDFs}. 

In figure~\ref{Fig:CascadeRDFs}b, we compare RDFs from the different methods and the~\citet{ZaichikAlipchenkov2009} model by scaling them with their respective $g_0$.
The curves are nearly identical.
Method A produces the highest RDF values, i.e. the most intermittent concentration distributions, and method C, as it is biased towards lower intermittency, results in the smallest asymptotic value (see table~\ref{Tab:AsymptoticValues}) for values of $g_0$.
Interestingly, $g_0$ for method C is the same as for the~\citet{ZaichikAlipchenkov2009} theory.
\citet{ZaichikAlipchenkov2009} use various approximations in their derivation. Among them, they model the turbulence by a Gaussian process which would underestimate the tails of their probability distributions and therefore underestimate intermittency. 

\begin{table}
  \begin{center}
\begin{tabular}{ccc}
    \hline \hline
    Simulation / Model                                    &  & $g_0$ \\ [3pt]
    \hline 
Cascade simulation -- Method A                &  & 13 \\
Cascade simulation -- Method B                &  & 5.9 \\
Cascade simulation -- Method C                &  & 3.9 \\
\citet{ZaichikAlipchenkov2009}, figure~3  &  & 3.9 \\
    \hline \hline
\end{tabular}
\caption{RDF asymptotic values, $g_0$, for $St_L^{-3/2} r/L \ll 1$ and $St_L \ll 1$ for the different cascade simulation methods, and the~\citet{ZaichikAlipchenkov2009} model.}
\label{Tab:AsymptoticValues}
\end{center}
\end{table}

\section{Discussion}
\label{Section:Discussion}











\subsection{The nature of scale-dependent particle concentration}
\label{Subsection:Discussion:Particles}

As described in section \ref{Section:Introduction}, the traditional explanation of clustering in terms of centrifugation of particles from eddies has been replaced with a somewhat more complicated and nuanced combination of physical processes \citep{BraggCollins2014I, BraggCollins2014II}. We believe that our results (figures~\ref{Fig:MultiplierBetas} (right) and~\ref{Fig:NominalModel}), stripped of their minor variations, represent a kind of universal curve (``U-curve") for $\beta(St_r)$ that can be interpreted in terms of these different processes acting on a particle of some $St$ over a range of scales $r$. 

In the $St_r  < 0.1$ regime, the effect is dominated by centrifugation, which weakens as $St_r$ decreases \citep[and others]{Chunetal2005}; thus $\beta$ increases (the multiplier PDF narrows) with decreasing $St_r$ in this regime. As $St_r$ increases beyond 1, the concentration effect is weakened by the decreasing sensitivity of particles to perturbations of any kind by eddies with timescales much shorter than their stopping times, so $\beta$ again increases. This effect, which could be thought of as an inertia impedance mismatch, has been described in terms of response functions \citep{Meek:1973qv, 1980A&A....85..316V, 1991A&A...242..286M, 2003Icar..164..127C, 2007A&A...466..413O}, but see~\citet{2010JFM...645..497B, 2013ApJ...776...12P, 2012MNRAS.426..784H} and~\citet{2014ApJ...797...59H} for other more recent and more sophisticated analyses. 

The strongest clustering effect is produced (the multiplier PDF has the lowest $\beta$) across perhaps one or two decades of eddy scale $r$ for a given $St$, centered on the combined parameter $St_r \sim 0.3$, presumably the regime where history effects in particle velocities play the dominant role. 
While our results support the idea that concentration is generically due to ``eddies on the scale of $\eta St_{\eta}^{3/2}$.." \citep{Irelandetal2015, 2007PhRvL..98h4502B}, we think a more refined description one could infer from the U-curve is that clustering is the {\it cumulative result of a history of interactions} with the flow of energy as it cascades over eddies ranging over two decades in size, driving particles ever deeper into a concentration ``attractor" even in the inertial range \citep{2009AnRFM..41..375T, BraggCollins2014I, BraggCollins2014II}. The other ``universal curve" of \citet{ZaichikAlipchenkov2003} (their figure 1) and their improved model~\citep[their figure 3]{ZaichikAlipchenkov2009} reproduced in figure~\ref{Fig:CascadeRDFs} also has this sense. That is, we see a parallel based on causality, between time-asymmetrical ``history effects" on particles of some $St$ as they are affected by energy flowing down the cascade through eddies of different scale, and a trajectory down one side and then back up the other side of our U-curve. Such a picture would lead the particle concentration as a function of spatial binning scale to increase sharply over some particular range of scales $r/L$ or $r/\eta$ related only to $St$ (the active range), and then remain constant towards smaller scales where eddy perturbations are felt only weakly because of, essentially, the poor impedance match with the particle stopping time. 

A natural prediction of this model is thus that, at infinite $Re$ where energy is available on all timescales, far from the dissipation range, and in the absence of complications such as gravitational settling or fragmentation limits, the maximum particle concentration should not only arise over a similar range of scales near $r \sim \eta St_{\eta}^{3/2}$ \citep{ZaichikAlipchenkov2003, ZaichikAlipchenkov2009, 2007PhRvL..98h4502B, Irelandetal2015}, but also should have a ``saturation" amplitude that is $St$-independent. Indeed this would seem to be the prediction of \citet[their figure 3]{ZaichikAlipchenkov2009}.  In current simulations \citep{Irelandetal2015, ZaichikAlipchenkov2009}, as well in simulations we have conducted using the cascade, the clustering of larger St particles ($St_L \gtrsim 0.03$) asymptotes at smaller values of the RDF than seen for smaller particles (figure~\ref{Fig:CascadeRDFs}). We expect this is because the scale at which the larger particles reach $St_r \sim 0.3$ is too close to the integral scale, so their potentially two-decade-wide range of interest, that our U-curve indicates is needed to reach a true asymptote, is truncated at large scales.

Within the dissipation range at $r < 20-30 \eta$, the energy spectrum of the flow changes as a result of the now-fixed eddy timescale $t_r =t_{\eta}$ \citep{Braggetal2015PRE}. Particles of $St_{\eta} \sim 1$ are now unique in that they do not experience the usual impedance mismatch with faster eddy forcing going to smaller spatial scales, so can continue to increase in concentration going to smaller scales. As noted by \citet{Braggetal2015PRE} the question remains as to whether there is any sort of rollover at $r \ll \eta$ in the RDF of $St_{\eta}$=1 particles, or whether an actual singularity would exist for point particles. In terrestrial applications, finite particle sizes comparable to $\eta$ preclude unlimited singular behavior; however, in protoplanetary nebula applications \citep{PanPadoan2010, 2013ApJ...776...12P, Johansenetal2015}, particle sizes of interest (submillimeter to dm) are orders of magnitude smaller than the Kolmogorov scale (km) so this is a question of significant interest. 

\citet{2007PhRvL..98h4502B} explicitly described the dissipation regime as characterized by an ``attractor" having fractal properties, and indeed multifractal properties were demonstrated by Hogan et al (1999) for clustering in this regime. It is known that cascades lead to fractal and multifractal spatial distributions \citep[and references therein]{1990PhRvA..41..894M, 1995JSP....78..311S}. We now suspect that the cascade of \citet{2007PhRvE..75e6305H}, in which the multiplier distributions do seem to obey level-{\it independent} scaling, were effectively {\it dissipation range cascades}.
Tests by \citet{2007PhRvE..75e6305H} showed good agreement between their level-independent cascade model and DNS.
However, the multiplier PDFs were determined at $3\eta$ and all their DNS results were for low-to-moderate $Re_{\lambda} < 140$ such that the integral scales were 14$\eta$, 24$\eta$, 45$\eta$, and 86$\eta$. At least the first three of these runs lie mostly within the dissipation range, where scaling does support level independence \citep{2007PhRvL..98h4502B, Braggetal2015PRE}. 
It might be worth exploring the use of dissipation range cascades further from the standpoint of modeling fractal structure or to study higher moments of the particle density PDF. Indeed \citet{BecChetrite2007} present what is, essentially, a cascade model that reproduces aspects of the particle concentration PDF. 

Moreover, to our knowledge, while fractal/multifractal behavior has been shown for particle clustering within the dissipation regime, either at scales of a few to tens of $\eta$ \citep{Hoganetal1999}, or scales smaller than $\eta$  \citep{Bec2003, Becetal2006}, no explicit study of this property in the inertial range has been done.  It would be of interest to find whether the inertial range cascade as described by the U-function (figure \ref{Fig:MultiplierBetas}), which is level-{\it dependent} but in a predictable way that is level-{\it independent}, would also produce such a distribution, when suitably scaled for St. This could be of use in modeling radiative transfer properties \citep{Shawetal2002}.

\subsection{Dissipation, enstrophy, and $Re$-dependence}
\label{Subsection:Discussion:DissipationAndEnstrophy}

As mentioned earlier in section~\ref{Subsection:Analysis:DissipationAndEnstrophy}, our dissipation multiplier PDFs have larger $\beta$ (are narrower) than the expected $\beta_\epsilon \sim 3$ for all binning sizes we could usefully study, reaching an asymptote of $\beta_\epsilon \sim 8$ at $r \lesssim L/86$. Based  on the very extensive inertial range manifested in figure~\ref{fig:inertialrange}~\citep{2010JFM...645..497B}, apparently extending up to $>2000\eta$, we had expected to find scale-free behavior in the dissipation multiplier PDF over most of this range. However, recalling figure~\ref{fig:chhabra}, especially as selected by larger $|q|$, which more strongly weight the structures where most dissipation occurs, the properties of dissipation are not invariant over as wide a range as is the second order velocity structure function that defines the inertial range in~\citet{2010JFM...645..497B}. 

\begin{figure}
   \centering
   \includegraphics[width=\linewidth]{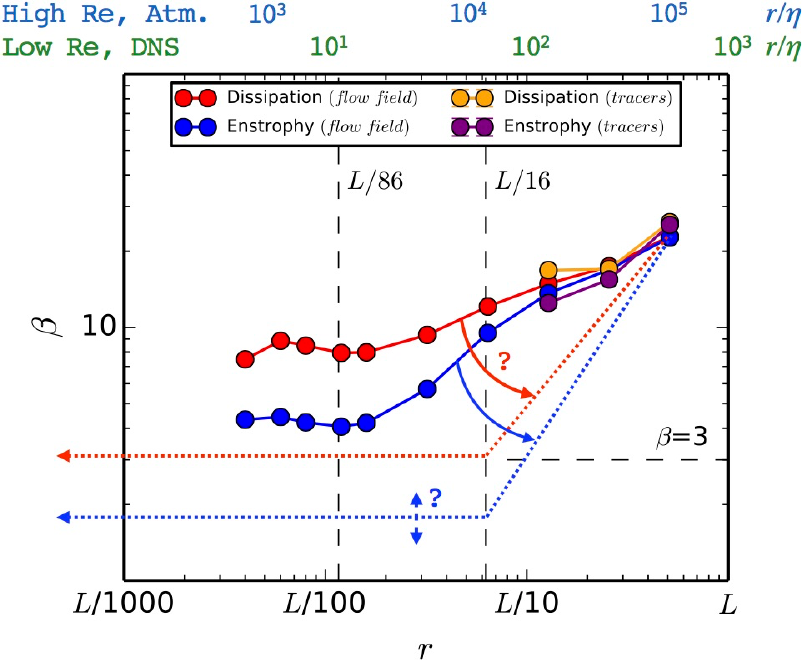}
   \caption{Hypothetical scaling behavior of dissipation and enstrophy: The solid red and blue symbols, as in figure~\ref{Fig:DissipationEnstrophyBeta}, are those we calculate from the numerical flow at $Re_{\lambda}=400$. Dotted lines are hypothetical values at very high $Re$. High-$Re$ atmospheric dissipation~\citep{1995JSP....78..311S} is scale-free at $\beta \approx 3$ at least between $L/86-L/860$; it is interesting that whatever changes affect the weighted quantities in figure~\ref{fig:chhabra} and the structure function in figure~\ref{fig:inertialrange}, the $\beta$ values for dissipation and enstropy do not seem to vary through the viscous (dissipation) subrange $r/\eta < 20$. It is plausible and expected that enstrophy will always be more intermittent than dissipation (have smaller $\beta$ values). The atmospheric (high-$Re$) value for the scale-free asymptote for dissipation is roughly $\beta$=3 (black dotted line). We hypothesize that $\beta$ values at high $Re$ follow trajectories similar to the red and blue dotted lines for dissipation and enstrophy respectively. That is, the observed behaviors (symbols) is the effect of an incompletely developed inertial range.
   \label{fig:hypothesis}}
\end{figure}

Figure~\ref{fig:hypothesis} summarizes the overall scale variation of $\beta$ for dissipation and enstrophy, showing the spatial scale both in terms of $\eta$ and $L$ following figure~\ref{fig:inertialrange}. At large scales, the PDF is narrow for both (large $\beta$) but widens with decreasing scale. At scales of $\sim 12\eta$ ($\sim L/86$) it asymptotes at a value which seems to remain scale-free to smaller scales. Our new asymptotic values do not agree with values ($\beta_{\cal E} \sim 10$ for enstrophy) found in~\citet{2007PhRvE..75e6305H}, or ($\beta_\epsilon \sim 3$ for dissipation) in~\citet{1995JSP....78..311S}.
It may be that the~\citet{2007PhRvE..75e6305H} $\beta_{\cal E}$ is more properly associated with the dissipation range, but the discrepancy in $\beta_\epsilon$ alone merits some discussion.

We hypothesize that at much higher $Re$ than we can study here, the scale-dependence of $\beta_{\epsilon}$ morphs in a fashion so as to be consistent with~\citet{1989PhRvA..40.5284C} and~\citet{1995JSP....78..311S}; that is, has a scale-free $\beta_{\epsilon} \sim 3$ for all scales less than at least 3000$\eta$ (based on~\citet{1989PhRvA..40.5284C}) and probably less than $L/16$ (based on~\citet{1989PhRvA..40.5284C} and~\citet{1989PhRvL..62.1327C}). At larger scales we expect $\beta$ must increase in some smooth fashion similar to ours, with an overall behavior schematically shown by the red dotted line in figure~\ref{fig:hypothesis}. Meanwhile, by the logic that enstrophy $\cal E$ is always more intermittent (has smaller $\beta$) than dissipation, we then also hypothesize that $\beta_{\cal E}$ varies as suggested by the blue dotted line. 

We suspect that our observed $\beta_\epsilon$ and $\beta_{\cal E}$ asymptote (for $r \lesssim L/86$) at larger values than would be true for much higher $Re$, because the viscous or dissipation range, which bounds the inertial range on its small-scale end and extends to 20-30$\eta$ in general, here impinges on the small-scale end of the nascent inertial range, and may prevent the dissipation and enstrophy from ever fully realizing their high-$Re$ intermittency.
In contrast, at high atmospheric $Re$, the large-scale onset of the inertial range, at $3000-10^4 \eta$ based on~\citet{1989PhRvA..40.5284C} and~\citet{1995JSP....78..311S}, is completely isolated from the viscous range, as indicated in figure~\ref{fig:inertialrange} by the blue axis labels on the upper horizontal axis.

The scale-dependence and asymptotic value of $\beta_{\cal E}$ is important, because the process of particle concentration may track the properties of $\cal E$ rather than those of $\epsilon$ based on the physics involved (section~\ref{Subsection:Discussion:Particles}). 
While it may be coincidental, in our DNS results, the $\beta$ for inertial particles minimizes at a value \textit{very close to our value for $\beta_{\cal E}$}, and considerably smaller (more intermittent) than our value of $\beta_\epsilon$. 
A secondary, related hypothesis is that the particle concentration multipliers may track the behavior of enstrophy (if velocities and accelerations are dominated by vorticity), and the minima seen in the collapsed curves of figure~\ref{Fig:MultiplierBetas}, which now never fall below 3.0, might drop to significantly lower values, making the particle concentration field more intermittent. For this reason, cascades developed using our \textit{current} collapsed $\beta(St_r)$ curves may underestimate the abundance of zones of high concentration at high $Re$ to some degree.  A better understanding of this $Re$-dependence will be needed to put cascade modeling of particle concentration on quantitatively solid ground. The original dataset of~\citet{1995JSP....78..311S} probably contains enough information to assess the validity of these hypotheses regarding dissipation, but not for enstrophy. 
 
\subsection{Speculations regarding the effect of higher $Re$-numbers}

As described in section~\ref{Section:NewCascade}, a level-dependent cascade can be described which captures the inertial range behavior for arbitrary $Re$ (at least to the point where mass loading starts to affect the physics, e.g., see~\citet{2007PhRvE..75e6305H}). This might be of use in modeling particle concentration in rain clouds, in the protoplanetary nebula, or in other applications at very high $Re$ where numerical simulations are impractical (such as spatial variations in microwave opacity).
There are several reasons to expect different behavior in the dissipation range.

Recalling, however, our numerical discrepancy with~\citet{1995JSP....78..311S} and others regarding the asymptotic value of $\beta$ for dissipation, we think the possibility of $Re$-dependence of even our collapsed $\beta(St_r)$ might imply that our results (and other inertial range results at low $Re$) underestimate effects of concentration, in the sense that at higher $Re$, the minimum in the universal curve would move to lower $\beta$ (more intermittency and higher concentrations).
One might speculate that the minimum $\beta$ for particles should track the $\beta$ for enstrophy instead of for dissipation.

In the application to planetesimal formation proposed by~\citet{2010Icar..208..518C}, it is necessary to create a joint PDF of particle concentration \textit{and} enstrophy. This motivates a better understanding of high $Re$ behavior of $\beta$ for both particles and enstrophy.

\section{Conclusions}
\label{Section:Conclusions}

Our results indicate that the multiplier PDFs for particle concentration (section~\ref{Subsection:Analysis:NonZeroStokesNumbers}), dissipation, and enstrophy (section~\ref{Subsection:Analysis:DissipationAndEnstrophy}) vary with scale, at least over the largest decade of spatial scales.
We also find that the multiplier PDFs for particle concentration have two components: a traditional ``$\beta$-function" component, and an exponential-tail component (section~\ref{Subsubsection:CompositePDFs}).
We find that the concentration multiplier $\beta$ values collapse to a \textit{scale}-independent universal curve when plotted against an appropriately scaled \textit{local Stokes number} $St_r = St (r/\eta)^{-2/3}$ (section~\ref{Subsubsection:MainResult}, in particular equation~(\ref{Eqn:LocalStokesNumber}) and figure~\ref{Fig:MultiplierBetas}), allowing the cascade model to be used for modeling higher $Re$ conditions not accessible to numerical simulation.

For dissipation, the ``$\beta \sim 3$" asymptotic behavior of~\citet{1995JSP....78..311S} in high $Re$ atmospheric flows probably appears at around $r \sim L/30$ or $L/40$, and remains constant to smaller scales, at least to $r \sim 20-30\eta$ where the dissipation range begins. 
In the present simulation, the integral scale and the dissipation range are not separated far enough for the dissipation $\beta_\epsilon$ to reach such low values, and instead it asymptotes for scales below $r \sim 20\eta$ to a value of $\beta_\epsilon \sim 8$.
Enstrophy, believed to be always more intermittent (smaller $\beta$) than dissipation, asymptotes in the DNS to a value of $\beta_{\cal E} \sim 3$.
In light of the connection between vorticity and the acceleration, centrifuging and concentrating of particles, it may not be surprising that this value coincides with the minimum $\beta$ value for particle concentration multipliers at the optimal $St_r$.

Given that dissipation, and presumably enstrophy, have not reached their scale-independent, asymptotic values seen in very high $Re$ atmospheric flows, it could be expected that our collapsed particle multiplier $\beta(St_r)$ curve is also $Re$ dependent.
Analyses of this sort for DNS of particle-laden flows at significantly higher Reynolds number are therefore highly desirable, as are measurements of enstrophy multipliers in very high Reynolds number flows such as atmospheric flows.

We have also found that the cascade model can be used to construct a spatial distribution of particle concentration, that can be carried to arbitrarily high Reynolds numbers, and has a very good resemblance to the analytical theory of~\citet{ZaichikAlipchenkov2009}. More work is needed to assess the asymptotic level of maximum concentration for particles of any size (which we find, as did \citet{ZaichikAlipchenkov2009}, is size invariant, in the infinite $Re$ limit).

\vspace{0.5cm}
\textbf{Acknowledgements:} The authors would like to thank Nic Brummell, Karim Shariff, Katepalli Sreenivasan, Federico Toschi and Alan Wray for insightful discussions on the topic of this paper, and Federico Toschi and Enrico Calzavarini for help accessing and using the DNS data.

We would like to dedicate this paper to Bob Hogan, who passed away in February 2012. Bob was responsible for all the previous computational work done on turbulent concentration by our group from 1992-2010, including multifractal behavior and our first cascades. He would be very interested in these new results, which explain a discrepancy that arose too late for him to resolve.

\appendix
\section{Three methods for computing the spatial distribution of concentrations}
\label{Appendix}
\numberwithin{equation}{section}

We here provide more details about the three methods for computing spatial concentration fields from multiplier distributions that we used in section~\ref{Subsection:RDF} for computing radial distribution functions.

Let us consider a cube of size $r/L=2^{-N/3}$ (cascade level $N$) having a concentration $C^{(N)}$, initially we start with a cascade level 0 cube having a unit concentration $C^{(0)}=1$, and divide it in half along each spatial direction.
This results in eight sub-cubes with half the linear size which correspond to a cascade level of $N+3$ (equation~\ref{Eqn:r_over_L}).
We denote the concentrations in these sub-cubes as $C_{ijk}^{(N+3)}$, where $i,j,k \in [1,2]$ are indices denoting the left (1) or right (2) sub-cube in the three spatial directions.
Since there are eight unknowns, we need eight equations to uniquely determine the concentrations. 

The first method -- \textit{method A} -- makes the following choice:
One constraint is given by the fact that the average of the concentrations over all eight sub-cubes is equal to the mean concentration, that is
\begin{equation}
  \sum_{i,j,k} \frac{1}{8} C^{(N+3)}_{ijk} = C^{(N)}.
   \label{Eqn:MethodA1} 
\end{equation}
We get seven more constraints by relating the concentration of neighboring sub-cubes to multipliers that are chosen randomly from the cascade model.
A possible choice are the combinations
\begin{eqnarray}
   (C^{(N+3)}_{111} + C^{(N+3)}_{211}) m_1^{(N+3)} & = & C^{(N+3)}_{111}, \nonumber \\
   (C^{(N+3)}_{112} + C^{(N+3)}_{212}) m_3^{(N+3)} & = & C^{(N+3)}_{112}, \nonumber \\
   (C^{(N+3)}_{111} + C^{(N+3)}_{121}) m_5^{(N+3)} & = & C^{(N+3)}_{111}, \nonumber \\
   (C^{(N+3)}_{112} + C^{(N+3)}_{122}) m_7^{(N+3)} & = & C^{(N+3)}_{112}, \nonumber \\
   (C^{(N+3)}_{121} + C^{(N+3)}_{221}) m_2^{(N+3)} & = & C^{(N+3)}_{221}, \nonumber \\
   (C^{(N+3)}_{122} + C^{(N+3)}_{222}) m_4^{(N+3)} & = & C^{(N+3)}_{122}, \nonumber \\
   C^{(N+3)}_{121} + C^{(N+3)}_{112}) m_6^{(N+3)} & = & C^{(N+3)}_{121},
   \label{Eqn:MethodA2}
\end{eqnarray}
where $m^{(N+3)}_i$ with $i \in [1,...,7]$ are seven random multiplier values at cascade level $N+3$.
Equations~(\ref{Eqn:MethodA1}) and (\ref{Eqn:MethodA2}) are linearly independent and can be solved directly.
Applying this procedure recursively to ever smaller cubes until some small cutoff length scale $(r/L)_\textrm{min}$ yields one statistical realization of the concentration field.

The two other methods are obtained by relating the concentrations in the two half-cubes, for each direction separately, through a random multiplier, that is
\begin{eqnarray}
    \sum_{j,k} C^{(N+3)}_{1ik}   =   2 C^{N} m_1^{(N+1)},& \,\,\,\,\, & \sum_{i,k} C^{(N+3)}_{i1k}   =   2 C^{N} m_2^{(N+1)},\nonumber \\
    \sum_{i,j} C^{(N+3)}_{ij1}  =  2 C^{N} m_3^{(N+1)}, & \,\,\,\,\, &
   \label{Eqn:MethodBC}
\end{eqnarray}
where $m^{(N+1)}_i$ with $i \in [1,...,3]$ are multipliers randomly drawn from the level $N+1$ cascade model.
Since these are only three equations, the linear system is underdetermined.

For \textit{method B} we choose one particular solution to equations~(\ref{Eqn:MethodBC}), namely
\begin{eqnarray}
   \frac{C^{(N+3)}_{111}}{C^{N}} & = & \left(\;\;\;\;\;\; m_1^{(N+1)}\right) \left(\;\;\;\;\;\; m_2^{(N+1)}\right) \left(\;\;\;\;\;\; m_3^{(N+1)}\right), \nonumber \\
   \frac{C^{(N+3)}_{211}}{C^{N}} & = & \left(1 - m_1^{(N+1)}\right) \left(\;\;\;\;\;\; m_2^{(N+1)}\right) \left(\;\;\;\;\;\; m_3^{(N+1)}\right), \nonumber \\
   \frac{C^{(N+3)}_{121}}{C^{N}} & = & \left(\;\;\;\;\;\; m_1^{(N+1)}\right) \left(1 - m_2^{(N+1)}\right) \left(\;\;\;\;\;\; m_3^{(N+1)}\right), \nonumber \\
   \frac{C^{(N+3)}_{221}}{C^{N}} & = & \left(1 - m_1^{(N+1)}\right) \left(1 - m_2^{(N+1)}\right) \left(\;\;\;\;\;\; m_3^{(N+1)}\right), \nonumber \\
   \frac{C^{(N+3)}_{112}}{C^{N}} & = & \left(\;\;\;\;\;\; m_1^{(N+1)}\right) \left(\;\;\;\;\;\; m_2^{(N+1)}\right) \left(1 - m_3^{(N+1)}\right), \nonumber \\
   \frac{C^{(N+3)}_{212}}{C^{N}} & = & \left(1 - m_1^{(N+1)}\right) \left(\;\;\;\;\;\; m_2^{(N+1)}\right) \left(1 - m_3^{(N+1)}\right), \nonumber \\
   \frac{C^{(N+3)}_{122}}{C^{N}} & = & \left(\;\;\;\;\;\; m_1^{(N+1)}\right) \left(1 - m_2^{(N+1)}\right) \left(1 - m_3^{(N+1)}\right), \nonumber \\
   \frac{C^{(N+3)}_{222}}{C^{N}} & = & \left(1 - m_1^{(N+1)}\right) \left(1 - m_2^{(N+1)}\right) \left(1 - m_3^{(N+1)}\right). \nonumber \\
   & \, &
   \label{Eqn:MethodB}
\end{eqnarray}
There is an ambiguity, of course, as to whether multiplier $m^{(N+1)}_i$ is applied to the ``left'' or ``right'' half of each box, but since the multipliers $m^{(N+1)}_i$ and $(1-m^{(N+1)}_i)$ have equal probability, either choice would be paired with the opposite given enough random samples (and we do average many random samples to construct each cascade).
The eight sub-cubes from each large box have the identical concentrations, just differently distributed, for each set of $m^{(N+1)}_i, i \in [1,...,3]$ whether $m^{(N+1)}_i$ goes to the left or right box in each case. 

For the third and final method, \textit{method C}, we use a least-square solver to determine the minimum-norm solution of (\ref{Eqn:MethodBC}), that is the solution that is closest to equally distributed concentrations. Clearly, this causes a bias towards the least intermittent spatial distribution. For strongly intermittent multipliers, this method can even lead to negative concentrations in one of the sub-cubes. Such solutions have to be discarded, and this further biases method C towards minimal intermittency.

\bibliographystyle{apsrev4-1}
\bibliography{bibliography}

\begin{thebibliography}{61}%
\makeatletter
\providecommand \@ifxundefined [1]{%
 \@ifx{#1\undefined}
}%
\providecommand \@ifnum [1]{%
 \ifnum #1\expandafter \@firstoftwo
 \else \expandafter \@secondoftwo
 \fi
}%
\providecommand \@ifx [1]{%
 \ifx #1\expandafter \@firstoftwo
 \else \expandafter \@secondoftwo
 \fi
}%
\providecommand \natexlab [1]{#1}%
\providecommand \enquote  [1]{``#1''}%
\providecommand \bibnamefont  [1]{#1}%
\providecommand \bibfnamefont [1]{#1}%
\providecommand \citenamefont [1]{#1}%
\providecommand \href@noop [0]{\@secondoftwo}%
\providecommand \href [0]{\begingroup \@sanitize@url \@href}%
\providecommand \@href[1]{\@@startlink{#1}\@@href}%
\providecommand \@@href[1]{\endgroup#1\@@endlink}%
\providecommand \@sanitize@url [0]{\catcode `\\12\catcode `\$12\catcode
  `\&12\catcode `\#12\catcode `\^12\catcode `\_12\catcode `\%12\relax}%
\providecommand \@@startlink[1]{}%
\providecommand \@@endlink[0]{}%
\providecommand \url  [0]{\begingroup\@sanitize@url \@url }%
\providecommand \@url [1]{\endgroup\@href {#1}{\urlprefix }}%
\providecommand \urlprefix  [0]{URL }%
\providecommand \Eprint [0]{\href }%
\providecommand \doibase [0]{http://dx.doi.org/}%
\providecommand \selectlanguage [0]{\@gobble}%
\providecommand \bibinfo  [0]{\@secondoftwo}%
\providecommand \bibfield  [0]{\@secondoftwo}%
\providecommand \translation [1]{[#1]}%
\providecommand \BibitemOpen [0]{}%
\providecommand \bibitemStop [0]{}%
\providecommand \bibitemNoStop [0]{.\EOS\space}%
\providecommand \EOS [0]{\spacefactor3000\relax}%
\providecommand \BibitemShut  [1]{\csname bibitem#1\endcsname}%
\let\auto@bib@innerbib\@empty
\bibitem [{\citenamefont {{Toschi}}\ and\ \citenamefont
  {{Bodenschatz}}(2009)}]{2009AnRFM..41..375T}%
  \BibitemOpen
  \bibfield  {author} {\bibinfo {author} {\bibfnamefont {F.}~\bibnamefont
  {{Toschi}}}\ and\ \bibinfo {author} {\bibfnamefont {E.}~\bibnamefont
  {{Bodenschatz}}},\ }\href {\doibase 10.1146/annurev.fluid.010908.165210}
  {\bibfield  {journal} {\bibinfo  {journal} {Annual Review of Fluid
  Mechanics}\ }\textbf {\bibinfo {volume} {41}},\ \bibinfo {pages} {375}
  (\bibinfo {year} {2009})}\BibitemShut {NoStop}%
\bibitem [{\citenamefont {{Zaichik}}\ and\ \citenamefont
  {{Alipchenkov}}(2009)}]{ZaichikAlipchenkov2009}%
  \BibitemOpen
  \bibfield  {author} {\bibinfo {author} {\bibfnamefont {L.~I.}\ \bibnamefont
  {{Zaichik}}}\ and\ \bibinfo {author} {\bibfnamefont {V.~M.}\ \bibnamefont
  {{Alipchenkov}}},\ }\href {\doibase 10.1088/1367-2630/11/10/103018}
  {\bibfield  {journal} {\bibinfo  {journal} {New Journal of Physics}\ }\textbf
  {\bibinfo {volume} {11}},\ \bibinfo {eid} {103018} (\bibinfo {year}
  {2009})}\BibitemShut {NoStop}%
\bibitem [{\citenamefont {{Bragg}}\ and\ \citenamefont
  {{Collins}}(2014{\natexlab{a}})}]{BraggCollins2014I}%
  \BibitemOpen
  \bibfield  {author} {\bibinfo {author} {\bibfnamefont {A.~D.}\ \bibnamefont
  {{Bragg}}}\ and\ \bibinfo {author} {\bibfnamefont {L.~R.}\ \bibnamefont
  {{Collins}}},\ }\href {\doibase 10.1088/1367-2630/16/5/055013} {\bibfield
  {journal} {\bibinfo  {journal} {New Journal of Physics}\ }\textbf {\bibinfo
  {volume} {16}},\ \bibinfo {eid} {055013} (\bibinfo {year}
  {2014}{\natexlab{a}})}\BibitemShut {NoStop}%
\bibitem [{\citenamefont {{Bragg}}\ and\ \citenamefont
  {{Collins}}(2014{\natexlab{b}})}]{BraggCollins2014II}%
  \BibitemOpen
  \bibfield  {author} {\bibinfo {author} {\bibfnamefont {A.~D.}\ \bibnamefont
  {{Bragg}}}\ and\ \bibinfo {author} {\bibfnamefont {L.~R.}\ \bibnamefont
  {{Collins}}},\ }\href {\doibase 10.1088/1367-2630/16/5/055014} {\bibfield
  {journal} {\bibinfo  {journal} {New Journal of Physics}\ }\textbf {\bibinfo
  {volume} {16}},\ \bibinfo {eid} {055014} (\bibinfo {year}
  {2014}{\natexlab{b}})}\BibitemShut {NoStop}%
\bibitem [{\citenamefont {{Gustavsson}}\ and\ \citenamefont
  {{Mehlig}}(2014)}]{GustavssonMehlig2014}%
  \BibitemOpen
  \bibfield  {author} {\bibinfo {author} {\bibfnamefont {K.}~\bibnamefont
  {{Gustavsson}}}\ and\ \bibinfo {author} {\bibfnamefont {B.}~\bibnamefont
  {{Mehlig}}},\ }\href@noop {} {\bibfield  {journal} {\bibinfo  {journal}
  {ArXiv e-prints; submitted to Ann. Rev. Flu. Mech.}\ } (\bibinfo {year}
  {2014})},\ \Eprint {http://arxiv.org/abs/1412.4374} {arXiv:1412.4374
  [physics.flu-dyn]} \BibitemShut {NoStop}%
\bibitem [{\citenamefont {{Maxey}}(1987)}]{Maxey1987}%
  \BibitemOpen
  \bibfield  {author} {\bibinfo {author} {\bibfnamefont {M.~R.}\ \bibnamefont
  {{Maxey}}},\ }\href {\doibase 10.1017/S0022112087000193} {\bibfield
  {journal} {\bibinfo  {journal} {Journal of Fluid Mechanics}\ }\textbf
  {\bibinfo {volume} {174}},\ \bibinfo {pages} {441} (\bibinfo {year}
  {1987})}\BibitemShut {NoStop}%
\bibitem [{\citenamefont {{Squires}}\ and\ \citenamefont
  {{Eaton}}(1991)}]{SquiresEaton1991}%
  \BibitemOpen
  \bibfield  {author} {\bibinfo {author} {\bibfnamefont {K.~D.}\ \bibnamefont
  {{Squires}}}\ and\ \bibinfo {author} {\bibfnamefont {J.~K.}\ \bibnamefont
  {{Eaton}}},\ }\href {\doibase 10.1063/1.858045} {\bibfield  {journal}
  {\bibinfo  {journal} {Physics of Fluids}\ }\textbf {\bibinfo {volume} {3}},\
  \bibinfo {pages} {1169} (\bibinfo {year} {1991})}\BibitemShut {NoStop}%
\bibitem [{\citenamefont {{Pan}}\ and\ \citenamefont
  {{Padoan}}(2010)}]{PanPadoan2010}%
  \BibitemOpen
  \bibfield  {author} {\bibinfo {author} {\bibfnamefont {L.}~\bibnamefont
  {{Pan}}}\ and\ \bibinfo {author} {\bibfnamefont {P.}~\bibnamefont
  {{Padoan}}},\ }\href {\doibase 10.1017/S0022112010002855} {\bibfield
  {journal} {\bibinfo  {journal} {Journal of Fluid Mechanics}\ }\textbf
  {\bibinfo {volume} {661}},\ \bibinfo {pages} {73} (\bibinfo {year} {2010})},\
  \Eprint {http://arxiv.org/abs/1005.2419} {arXiv:1005.2419 [physics.flu-dyn]}
  \BibitemShut {NoStop}%
\bibitem [{\citenamefont {{Pan}}\ and\ \citenamefont
  {{Padoan}}(2013)}]{2013ApJ...776...12P}%
  \BibitemOpen
  \bibfield  {author} {\bibinfo {author} {\bibfnamefont {L.}~\bibnamefont
  {{Pan}}}\ and\ \bibinfo {author} {\bibfnamefont {P.}~\bibnamefont
  {{Padoan}}},\ }\href {\doibase 10.1088/0004-637X/776/1/12} {\bibfield
  {journal} {\bibinfo  {journal} {\apj}\ }\textbf {\bibinfo {volume} {776}},\
  \bibinfo {eid} {12} (\bibinfo {year} {2013})},\ \Eprint
  {http://arxiv.org/abs/1305.0307} {arXiv:1305.0307 [astro-ph.EP]} \BibitemShut
  {NoStop}%
\bibitem [{\citenamefont {{Bragg}}\ \emph
  {et~al.}(2015{\natexlab{a}})\citenamefont {{Bragg}}, \citenamefont
  {{Ireland}},\ and\ \citenamefont {{Collins}}}]{Braggetal2015JFM}%
  \BibitemOpen
  \bibfield  {author} {\bibinfo {author} {\bibfnamefont {A.~D.}\ \bibnamefont
  {{Bragg}}}, \bibinfo {author} {\bibfnamefont {P.~J.}\ \bibnamefont
  {{Ireland}}}, \ and\ \bibinfo {author} {\bibfnamefont {L.~R.}\ \bibnamefont
  {{Collins}}},\ }\href {\doibase 10.1017/jfm.2015.474} {\bibfield  {journal}
  {\bibinfo  {journal} {Journal of Fluid Mechanics}\ }\textbf {\bibinfo
  {volume} {780}},\ \bibinfo {pages} {327} (\bibinfo {year}
  {2015}{\natexlab{a}})},\ \Eprint {http://arxiv.org/abs/1501.03842}
  {arXiv:1501.03842 [physics.flu-dyn]} \BibitemShut {NoStop}%
\bibitem [{\citenamefont {{Zaichik}}\ and\ \citenamefont
  {{Alipchenkov}}(2003)}]{ZaichikAlipchenkov2003}%
  \BibitemOpen
  \bibfield  {author} {\bibinfo {author} {\bibfnamefont {L.~I.}\ \bibnamefont
  {{Zaichik}}}\ and\ \bibinfo {author} {\bibfnamefont {V.~M.}\ \bibnamefont
  {{Alipchenkov}}},\ }\href {\doibase 10.1063/1.1569485} {\bibfield  {journal}
  {\bibinfo  {journal} {Physics of Fluids}\ }\textbf {\bibinfo {volume} {15}},\
  \bibinfo {pages} {1776} (\bibinfo {year} {2003})}\BibitemShut {NoStop}%
\bibitem [{\citenamefont {{Bec}}\ \emph
  {et~al.}(2010{\natexlab{a}})\citenamefont {{Bec}}, \citenamefont
  {{Biferale}}, \citenamefont {{Cencini}}, \citenamefont {{Lanotte}},\ and\
  \citenamefont {{Toschi}}}]{2010JFM...646..527B}%
  \BibitemOpen
  \bibfield  {author} {\bibinfo {author} {\bibfnamefont {J.}~\bibnamefont
  {{Bec}}}, \bibinfo {author} {\bibfnamefont {L.}~\bibnamefont {{Biferale}}},
  \bibinfo {author} {\bibfnamefont {M.}~\bibnamefont {{Cencini}}}, \bibinfo
  {author} {\bibfnamefont {A.~S.}\ \bibnamefont {{Lanotte}}}, \ and\ \bibinfo
  {author} {\bibfnamefont {F.}~\bibnamefont {{Toschi}}},\ }\href {\doibase
  10.1017/S0022112010000029} {\bibfield  {journal} {\bibinfo  {journal}
  {Journal of Fluid Mechanics}\ }\textbf {\bibinfo {volume} {646}},\ \bibinfo
  {pages} {527} (\bibinfo {year} {2010}{\natexlab{a}})}\BibitemShut {NoStop}%
\bibitem [{\citenamefont {{Ireland}}\ \emph {et~al.}(2015)\citenamefont
  {{Ireland}}, \citenamefont {{Bragg}},\ and\ \citenamefont
  {{Collins}}}]{Irelandetal2015}%
  \BibitemOpen
  \bibfield  {author} {\bibinfo {author} {\bibfnamefont {P.~J.}\ \bibnamefont
  {{Ireland}}}, \bibinfo {author} {\bibfnamefont {A.~D.}\ \bibnamefont
  {{Bragg}}}, \ and\ \bibinfo {author} {\bibfnamefont {L.~R.}\ \bibnamefont
  {{Collins}}},\ }\href@noop {} {\bibfield  {journal} {\bibinfo  {journal}
  {ArXiv e-prints}\ } (\bibinfo {year} {2015})},\ \Eprint
  {http://arxiv.org/abs/1507.07026} {arXiv:1507.07026 [physics.flu-dyn]}
  \BibitemShut {NoStop}%
\bibitem [{\citenamefont {{Pan}}\ and\ \citenamefont
  {{Padoan}}(2014)}]{PanPadoan2014}%
  \BibitemOpen
  \bibfield  {author} {\bibinfo {author} {\bibfnamefont {L.}~\bibnamefont
  {{Pan}}}\ and\ \bibinfo {author} {\bibfnamefont {P.}~\bibnamefont
  {{Padoan}}},\ }\href {\doibase 10.1088/0004-637X/797/2/101} {\bibfield
  {journal} {\bibinfo  {journal} {\apj}\ }\textbf {\bibinfo {volume} {797}},\
  \bibinfo {eid} {101} (\bibinfo {year} {2014})},\ \Eprint
  {http://arxiv.org/abs/1410.1989} {arXiv:1410.1989 [astro-ph.EP]} \BibitemShut
  {NoStop}%
\bibitem [{\citenamefont {{Pan}}\ and\ \citenamefont
  {{Padoan}}(2015)}]{PanPadoan2015}%
  \BibitemOpen
  \bibfield  {author} {\bibinfo {author} {\bibfnamefont {L.}~\bibnamefont
  {{Pan}}}\ and\ \bibinfo {author} {\bibfnamefont {P.}~\bibnamefont
  {{Padoan}}},\ }\href {\doibase 10.1088/0004-637X/812/1/10} {\bibfield
  {journal} {\bibinfo  {journal} {\apj}\ }\textbf {\bibinfo {volume} {812}},\
  \bibinfo {eid} {10} (\bibinfo {year} {2015})}\BibitemShut {NoStop}%
\bibitem [{\citenamefont {{Shaw}}\ \emph {et~al.}(2002)\citenamefont {{Shaw}},
  \citenamefont {{Kostinski}},\ and\ \citenamefont
  {{Lanterman}}}]{Shawetal2002}%
  \BibitemOpen
  \bibfield  {author} {\bibinfo {author} {\bibfnamefont {R.~A.}\ \bibnamefont
  {{Shaw}}}, \bibinfo {author} {\bibfnamefont {A.~B.}\ \bibnamefont
  {{Kostinski}}}, \ and\ \bibinfo {author} {\bibfnamefont {D.~D.}\ \bibnamefont
  {{Lanterman}}},\ }\href {\doibase 10.1016/S0022-4073(01)00287-4} {\bibfield
  {journal} {\bibinfo  {journal} {\jqsrt}\ }\textbf {\bibinfo {volume} {75}},\
  \bibinfo {pages} {13} (\bibinfo {year} {2002})}\BibitemShut {NoStop}%
\bibitem [{\citenamefont {{Cuzzi}}\ \emph {et~al.}(2008)\citenamefont
  {{Cuzzi}}, \citenamefont {{Hogan}},\ and\ \citenamefont
  {{Shariff}}}]{2008ApJ...687.1432C}%
  \BibitemOpen
  \bibfield  {author} {\bibinfo {author} {\bibfnamefont {J.~N.}\ \bibnamefont
  {{Cuzzi}}}, \bibinfo {author} {\bibfnamefont {R.~C.}\ \bibnamefont
  {{Hogan}}}, \ and\ \bibinfo {author} {\bibfnamefont {K.}~\bibnamefont
  {{Shariff}}},\ }\href {\doibase 10.1086/591239} {\bibfield  {journal}
  {\bibinfo  {journal} {\apj}\ }\textbf {\bibinfo {volume} {687}},\ \bibinfo
  {pages} {1432} (\bibinfo {year} {2008})},\ \Eprint
  {http://arxiv.org/abs/0804.3526} {arXiv:0804.3526} \BibitemShut {NoStop}%
\bibitem [{\citenamefont {{Cuzzi}}\ \emph {et~al.}(2010)\citenamefont
  {{Cuzzi}}, \citenamefont {{Hogan}},\ and\ \citenamefont
  {{Bottke}}}]{2010Icar..208..518C}%
  \BibitemOpen
  \bibfield  {author} {\bibinfo {author} {\bibfnamefont {J.~N.}\ \bibnamefont
  {{Cuzzi}}}, \bibinfo {author} {\bibfnamefont {R.~C.}\ \bibnamefont
  {{Hogan}}}, \ and\ \bibinfo {author} {\bibfnamefont {W.~F.}\ \bibnamefont
  {{Bottke}}},\ }\href {\doibase 10.1016/j.icarus.2010.03.005} {\bibfield
  {journal} {\bibinfo  {journal} {Icarus}\ }\textbf {\bibinfo {volume} {208}},\
  \bibinfo {pages} {518} (\bibinfo {year} {2010})},\ \Eprint
  {http://arxiv.org/abs/1004.0270} {arXiv:1004.0270 [astro-ph.EP]} \BibitemShut
  {NoStop}%
\bibitem [{\citenamefont {{Chambers}}(2010)}]{Chambers2010}%
  \BibitemOpen
  \bibfield  {author} {\bibinfo {author} {\bibfnamefont {J.~E.}\ \bibnamefont
  {{Chambers}}},\ }\href {\doibase 10.1016/j.icarus.2010.03.004} {\bibfield
  {journal} {\bibinfo  {journal} {\icarus}\ }\textbf {\bibinfo {volume}
  {208}},\ \bibinfo {pages} {505} (\bibinfo {year} {2010})}\BibitemShut
  {NoStop}%
\bibitem [{\citenamefont {Pan}\ \emph {et~al.}(2011)\citenamefont {Pan},
  \citenamefont {Padoan}, \citenamefont {Scalo}, \citenamefont {Kritsuk},\ and\
  \citenamefont {Norman}}]{0004-637X-740-1-6}%
  \BibitemOpen
  \bibfield  {author} {\bibinfo {author} {\bibfnamefont {L.}~\bibnamefont
  {Pan}}, \bibinfo {author} {\bibfnamefont {P.}~\bibnamefont {Padoan}},
  \bibinfo {author} {\bibfnamefont {J.}~\bibnamefont {Scalo}}, \bibinfo
  {author} {\bibfnamefont {A.~G.}\ \bibnamefont {Kritsuk}}, \ and\ \bibinfo
  {author} {\bibfnamefont {M.~L.}\ \bibnamefont {Norman}},\ }\href
  {http://stacks.iop.org/0004-637X/740/i=1/a=6} {\bibfield  {journal} {\bibinfo
   {journal} {The Astrophysical Journal}\ }\textbf {\bibinfo {volume} {740}},\
  \bibinfo {pages} {6} (\bibinfo {year} {2011})}\BibitemShut {NoStop}%
\bibitem [{\citenamefont {{Hopkins}}(2014)}]{2014ApJ...797...59H}%
  \BibitemOpen
  \bibfield  {author} {\bibinfo {author} {\bibfnamefont {P.~F.}\ \bibnamefont
  {{Hopkins}}},\ }\href {\doibase 10.1088/0004-637X/797/1/59} {\bibfield
  {journal} {\bibinfo  {journal} {\apj}\ }\textbf {\bibinfo {volume} {797}},\
  \bibinfo {eid} {59} (\bibinfo {year} {2014})},\ \Eprint
  {http://arxiv.org/abs/1406.5509} {arXiv:1406.5509} \BibitemShut {NoStop}%
\bibitem [{\citenamefont {{Johansen}}\ \emph {et~al.}(2015)\citenamefont
  {{Johansen}}, \citenamefont {{Jacquet}}, \citenamefont {{Cuzzi}},
  \citenamefont {{Morbidelli}},\ and\ \citenamefont
  {{Gounelle}}}]{Johansenetal2015}%
  \BibitemOpen
  \bibfield  {author} {\bibinfo {author} {\bibfnamefont {A.}~\bibnamefont
  {{Johansen}}}, \bibinfo {author} {\bibfnamefont {E.}~\bibnamefont
  {{Jacquet}}}, \bibinfo {author} {\bibfnamefont {J.~N.}\ \bibnamefont
  {{Cuzzi}}}, \bibinfo {author} {\bibfnamefont {A.}~\bibnamefont
  {{Morbidelli}}}, \ and\ \bibinfo {author} {\bibfnamefont {M.}~\bibnamefont
  {{Gounelle}}},\ }\enquote {\bibinfo {title} {{New Paradigms for Asteroid
  Formation}},}\ in\ \href {\doibase 10.2458/azu_uapress_9780816530595-ch025}
  {\emph {\bibinfo {booktitle} {Asteroids IV}}},\ \bibinfo {editor} {edited by\
  \bibinfo {editor} {\bibfnamefont {P.}~\bibnamefont {{Michel}}}, \bibinfo
  {editor} {\bibfnamefont {F.~E.}\ \bibnamefont {{DeMeo}}}, \ and\ \bibinfo
  {editor} {\bibfnamefont {W.~F.}\ \bibnamefont {{Bottke}}}}\ (\bibinfo {year}
  {2015})\ pp.\ \bibinfo {pages} {471--492}\BibitemShut {NoStop}%
\bibitem [{\citenamefont {{Hopkins}}\ and\ \citenamefont
  {{Lee}}(2015)}]{2015arXiv151002477H}%
  \BibitemOpen
  \bibfield  {author} {\bibinfo {author} {\bibfnamefont {P.~F.}\ \bibnamefont
  {{Hopkins}}}\ and\ \bibinfo {author} {\bibfnamefont {H.}~\bibnamefont
  {{Lee}}},\ }\href@noop {} {\bibfield  {journal} {\bibinfo  {journal} {ArXiv
  e-prints}\ } (\bibinfo {year} {2015})},\ \Eprint
  {http://arxiv.org/abs/1510.02477} {arXiv:1510.02477} \BibitemShut {NoStop}%
\bibitem [{\citenamefont {{Bec}}\ \emph {et~al.}(2007)\citenamefont {{Bec}},
  \citenamefont {{Biferale}}, \citenamefont {{Cencini}}, \citenamefont
  {{Lanotte}}, \citenamefont {{Musacchio}},\ and\ \citenamefont
  {{Toschi}}}]{2007PhRvL..98h4502B}%
  \BibitemOpen
  \bibfield  {author} {\bibinfo {author} {\bibfnamefont {J.}~\bibnamefont
  {{Bec}}}, \bibinfo {author} {\bibfnamefont {L.}~\bibnamefont {{Biferale}}},
  \bibinfo {author} {\bibfnamefont {M.}~\bibnamefont {{Cencini}}}, \bibinfo
  {author} {\bibfnamefont {A.}~\bibnamefont {{Lanotte}}}, \bibinfo {author}
  {\bibfnamefont {S.}~\bibnamefont {{Musacchio}}}, \ and\ \bibinfo {author}
  {\bibfnamefont {F.}~\bibnamefont {{Toschi}}},\ }\href {\doibase
  10.1103/PhysRevLett.98.084502} {\bibfield  {journal} {\bibinfo  {journal}
  {Physical Review Letters}\ }\textbf {\bibinfo {volume} {98}},\ \bibinfo {eid}
  {084502} (\bibinfo {year} {2007})},\ \Eprint
  {http://arxiv.org/abs/nlin/0608045} {nlin/0608045} \BibitemShut {NoStop}%
\bibitem [{\citenamefont {{Bragg}}\ \emph
  {et~al.}(2015{\natexlab{b}})\citenamefont {{Bragg}}, \citenamefont
  {{Ireland}},\ and\ \citenamefont {{Collins}}}]{Braggetal2015PRE}%
  \BibitemOpen
  \bibfield  {author} {\bibinfo {author} {\bibfnamefont {A.~D.}\ \bibnamefont
  {{Bragg}}}, \bibinfo {author} {\bibfnamefont {P.~J.}\ \bibnamefont
  {{Ireland}}}, \ and\ \bibinfo {author} {\bibfnamefont {L.~R.}\ \bibnamefont
  {{Collins}}},\ }\href {\doibase 10.1103/PhysRevE.92.023029} {\bibfield
  {journal} {\bibinfo  {journal} {\pre}\ }\textbf {\bibinfo {volume} {92}},\
  \bibinfo {eid} {023029} (\bibinfo {year} {2015}{\natexlab{b}})},\ \Eprint
  {http://arxiv.org/abs/1411.7422} {arXiv:1411.7422 [physics.flu-dyn]}
  \BibitemShut {NoStop}%
\bibitem [{\citenamefont {{Meneveau}}\ and\ \citenamefont
  {{Sreenivasan}}(1987{\natexlab{a}})}]{1987PhRvL..59.1424M}%
  \BibitemOpen
  \bibfield  {author} {\bibinfo {author} {\bibfnamefont {C.}~\bibnamefont
  {{Meneveau}}}\ and\ \bibinfo {author} {\bibfnamefont {K.~R.}\ \bibnamefont
  {{Sreenivasan}}},\ }\href {\doibase 10.1103/PhysRevLett.59.1424} {\bibfield
  {journal} {\bibinfo  {journal} {Physical Review Letters}\ }\textbf {\bibinfo
  {volume} {59}},\ \bibinfo {pages} {1424} (\bibinfo {year}
  {1987}{\natexlab{a}})}\BibitemShut {NoStop}%
\bibitem [{\citenamefont {{Meneveau}}\ and\ \citenamefont
  {{Sreenivasan}}(1991)}]{1991JFM...224..429M}%
  \BibitemOpen
  \bibfield  {author} {\bibinfo {author} {\bibfnamefont {C.}~\bibnamefont
  {{Meneveau}}}\ and\ \bibinfo {author} {\bibfnamefont {K.~R.}\ \bibnamefont
  {{Sreenivasan}}},\ }\href {\doibase 10.1017/S0022112091001830} {\bibfield
  {journal} {\bibinfo  {journal} {Journal of Fluid Mechanics}\ }\textbf
  {\bibinfo {volume} {224}},\ \bibinfo {pages} {429} (\bibinfo {year}
  {1991})}\BibitemShut {NoStop}%
\bibitem [{\citenamefont {{Chhabra}}\ \emph {et~al.}(1989)\citenamefont
  {{Chhabra}}, \citenamefont {{Meneveau}}, \citenamefont {{Jensen}},\ and\
  \citenamefont {{Sreenivasan}}}]{1989PhRvA..40.5284C}%
  \BibitemOpen
  \bibfield  {author} {\bibinfo {author} {\bibfnamefont {A.~B.}\ \bibnamefont
  {{Chhabra}}}, \bibinfo {author} {\bibfnamefont {C.}~\bibnamefont
  {{Meneveau}}}, \bibinfo {author} {\bibfnamefont {R.~V.}\ \bibnamefont
  {{Jensen}}}, \ and\ \bibinfo {author} {\bibfnamefont {K.~R.}\ \bibnamefont
  {{Sreenivasan}}},\ }\href {\doibase 10.1103/PhysRevA.40.5284} {\bibfield
  {journal} {\bibinfo  {journal} {\pra}\ }\textbf {\bibinfo {volume} {40}},\
  \bibinfo {pages} {5284} (\bibinfo {year} {1989})}\BibitemShut {NoStop}%
\bibitem [{\citenamefont {{Chhabra}}\ and\ \citenamefont
  {{Sreenivasan}}(1992)}]{1992PhRvL..68.2762C}%
  \BibitemOpen
  \bibfield  {author} {\bibinfo {author} {\bibfnamefont {A.~B.}\ \bibnamefont
  {{Chhabra}}}\ and\ \bibinfo {author} {\bibfnamefont {K.~R.}\ \bibnamefont
  {{Sreenivasan}}},\ }\href {\doibase 10.1103/PhysRevLett.68.2762} {\bibfield
  {journal} {\bibinfo  {journal} {Physical Review Letters}\ }\textbf {\bibinfo
  {volume} {68}},\ \bibinfo {pages} {2762} (\bibinfo {year}
  {1992})}\BibitemShut {NoStop}%
\bibitem [{\citenamefont {Sreenivasan}\ and\ \citenamefont
  {Stolovitzky}(1994)}]{1994ActaMechSup4..113S}%
  \BibitemOpen
  \bibfield  {author} {\bibinfo {author} {\bibfnamefont {K.}~\bibnamefont
  {Sreenivasan}}\ and\ \bibinfo {author} {\bibfnamefont {G.}~\bibnamefont
  {Stolovitzky}},\ }\href {\doibase 10.1007/978-3-7091-9310-5_13} {\bibfield
  {journal} {\bibinfo  {journal} {Acta Mechanica (Suppl.)}\ }\textbf {\bibinfo
  {volume} {4}},\ \bibinfo {pages} {113} (\bibinfo {year} {1994})}\BibitemShut
  {NoStop}%
\bibitem [{\citenamefont {{Sreenivasan}}\ and\ \citenamefont
  {{Stolovitzky}}(1995)}]{1995JSP....78..311S}%
  \BibitemOpen
  \bibfield  {author} {\bibinfo {author} {\bibfnamefont {K.~R.}\ \bibnamefont
  {{Sreenivasan}}}\ and\ \bibinfo {author} {\bibfnamefont {G.}~\bibnamefont
  {{Stolovitzky}}},\ }\href {\doibase 10.1007/BF02183351} {\bibfield  {journal}
  {\bibinfo  {journal} {Journal of Statistical Physics}\ }\textbf {\bibinfo
  {volume} {78}},\ \bibinfo {pages} {311} (\bibinfo {year} {1995})}\BibitemShut
  {NoStop}%
\bibitem [{\citenamefont {{Hogan}}\ and\ \citenamefont
  {{Cuzzi}}(2007)}]{2007PhRvE..75e6305H}%
  \BibitemOpen
  \bibfield  {author} {\bibinfo {author} {\bibfnamefont {R.~C.}\ \bibnamefont
  {{Hogan}}}\ and\ \bibinfo {author} {\bibfnamefont {J.~N.}\ \bibnamefont
  {{Cuzzi}}},\ }\href {\doibase 10.1103/PhysRevE.75.056305} {\bibfield
  {journal} {\bibinfo  {journal} {\pre}\ }\textbf {\bibinfo {volume} {75}},\
  \bibinfo {eid} {056305} (\bibinfo {year} {2007})},\ \Eprint
  {http://arxiv.org/abs/0704.1810} {arXiv:0704.1810} \BibitemShut {NoStop}%
\bibitem [{\citenamefont {{Tennekes}}\ and\ \citenamefont
  {{Lumley}}(1972)}]{1972fct..book.....T}%
  \BibitemOpen
  \bibfield  {author} {\bibinfo {author} {\bibfnamefont {H.}~\bibnamefont
  {{Tennekes}}}\ and\ \bibinfo {author} {\bibfnamefont {J.~L.}\ \bibnamefont
  {{Lumley}}},\ }\href@noop {} {\emph {\bibinfo {title} {First Course in
  Turbulence}}}\ (\bibinfo  {publisher} {MIT Press, Cambridge},\ \bibinfo
  {year} {1972})\BibitemShut {NoStop}%
\bibitem [{\citenamefont {{Frisch}}(1995)}]{1995tlan.book.....F}%
  \BibitemOpen
  \bibfield  {author} {\bibinfo {author} {\bibfnamefont {U.}~\bibnamefont
  {{Frisch}}},\ }\href@noop {} {\emph {\bibinfo {title} {Turbulence.~The legacy
  of A.~N.~Kolmogorov.}}}\ (\bibinfo  {publisher} {Cambridge University Press,
  Cambridge (UK)},\ \bibinfo {year} {1995})\BibitemShut {NoStop}%
\bibitem [{\citenamefont {Kholmyanskiy}(1973)}]{Kholmyanskiy:1973tg}%
  \BibitemOpen
  \bibfield  {author} {\bibinfo {author} {\bibfnamefont {M.~Z.}\ \bibnamefont
  {Kholmyanskiy}},\ }\href@noop {} {\bibfield  {journal} {\bibinfo  {journal}
  {Izv. Atmos. Ocean. Phys.}\ }\textbf {\bibinfo {volume} {9}},\ \bibinfo
  {pages} {801} (\bibinfo {year} {1973})}\BibitemShut {NoStop}%
\bibitem [{\citenamefont {{van Atta}}\ and\ \citenamefont
  {{Yeh}}(1975)}]{1975JFM....71..417V}%
  \BibitemOpen
  \bibfield  {author} {\bibinfo {author} {\bibfnamefont {C.~W.}\ \bibnamefont
  {{van Atta}}}\ and\ \bibinfo {author} {\bibfnamefont {T.~T.}\ \bibnamefont
  {{Yeh}}},\ }\href {\doibase 10.1017/S0022112075002650} {\bibfield  {journal}
  {\bibinfo  {journal} {Journal of Fluid Mechanics}\ }\textbf {\bibinfo
  {volume} {71}},\ \bibinfo {pages} {417} (\bibinfo {year} {1975})}\BibitemShut
  {NoStop}%
\bibitem [{\citenamefont {{Meneveau}}\ and\ \citenamefont
  {{Sreenivasan}}(1987{\natexlab{b}})}]{1987NuPhS...2...49M}%
  \BibitemOpen
  \bibfield  {author} {\bibinfo {author} {\bibfnamefont {C.}~\bibnamefont
  {{Meneveau}}}\ and\ \bibinfo {author} {\bibfnamefont {K.~R.}\ \bibnamefont
  {{Sreenivasan}}},\ }\href {\doibase 10.1016/0920-5632(87)90008-9} {\bibfield
  {journal} {\bibinfo  {journal} {Nuclear Physics B Proceedings Supplements}\
  }\textbf {\bibinfo {volume} {2}},\ \bibinfo {pages} {49} (\bibinfo {year}
  {1987}{\natexlab{b}})}\BibitemShut {NoStop}%
\bibitem [{Note1()}]{Note1}%
  \BibitemOpen
  \bibinfo {note} {These results were cited and reanalyzed by~\protect \citet
  {1995JSP....78..311S}, however the reference to the basic data given
  by~\protect \citet {1995JSP....78..311S} is confused with an interpretive
  article by~\protect \citet {1989PhRvL..62.1327C}, who themselves
  cite~\protect \citet {1987PhRvL..59.1424M} and~\protect \citet [at the time
  unpublished]{1989PhRvA..40.5284C} for discussions and analysis of the basic
  data.}\BibitemShut {Stop}%
\bibitem [{\citenamefont {Hunt}\ and\ \citenamefont
  {Morrison}(2000)}]{Hunt:2000kb}%
  \BibitemOpen
  \bibfield  {author} {\bibinfo {author} {\bibfnamefont {J.~C.~R.}\
  \bibnamefont {Hunt}}\ and\ \bibinfo {author} {\bibfnamefont {J.~F.}\
  \bibnamefont {Morrison}},\ }\href@noop {} {\bibfield  {journal} {\bibinfo
  {journal} {Eur. J. Mech. B - Fluids}\ }\textbf {\bibinfo {volume} {19}},\
  \bibinfo {pages} {673} (\bibinfo {year} {2000})}\BibitemShut {NoStop}%
\bibitem [{\citenamefont {{Arn{\`e}odo}}\ \emph {et~al.}(2008)\citenamefont
  {{Arn{\`e}odo}}, \citenamefont {{Benzi}}, \citenamefont {{Berg}},
  \citenamefont {{Biferale}}, \citenamefont {{Bodenschatz}}, \citenamefont
  {{Busse}}, \citenamefont {{Calzavarini}}, \citenamefont {{Castaing}},
  \citenamefont {{Cencini}}, \citenamefont {{Chevillard}}, \citenamefont
  {{Fisher}}, \citenamefont {{Grauer}}, \citenamefont {{Homann}}, \citenamefont
  {{Lamb}}, \citenamefont {{Lanotte}}, \citenamefont {{L{\'e}v{\`e}que}},
  \citenamefont {{L{\"u}thi}}, \citenamefont {{Mann}}, \citenamefont
  {{Mordant}}, \citenamefont {{M{\"u}ller}}, \citenamefont {{Ott}},
  \citenamefont {{Ouellette}}, \citenamefont {{Pinton}}, \citenamefont
  {{Pope}}, \citenamefont {{Roux}}, \citenamefont {{Toschi}}, \citenamefont
  {{Xu}},\ and\ \citenamefont {{Yeung}}}]{2008PhRvL.100y4504A}%
  \BibitemOpen
  \bibfield  {author} {\bibinfo {author} {\bibfnamefont {A.}~\bibnamefont
  {{Arn{\`e}odo}}}, \bibinfo {author} {\bibfnamefont {R.}~\bibnamefont
  {{Benzi}}}, \bibinfo {author} {\bibfnamefont {J.}~\bibnamefont {{Berg}}},
  \bibinfo {author} {\bibfnamefont {L.}~\bibnamefont {{Biferale}}}, \bibinfo
  {author} {\bibfnamefont {E.}~\bibnamefont {{Bodenschatz}}}, \bibinfo {author}
  {\bibfnamefont {A.}~\bibnamefont {{Busse}}}, \bibinfo {author} {\bibfnamefont
  {E.}~\bibnamefont {{Calzavarini}}}, \bibinfo {author} {\bibfnamefont
  {B.}~\bibnamefont {{Castaing}}}, \bibinfo {author} {\bibfnamefont
  {M.}~\bibnamefont {{Cencini}}}, \bibinfo {author} {\bibfnamefont
  {L.}~\bibnamefont {{Chevillard}}}, \bibinfo {author} {\bibfnamefont {R.~T.}\
  \bibnamefont {{Fisher}}}, \bibinfo {author} {\bibfnamefont {R.}~\bibnamefont
  {{Grauer}}}, \bibinfo {author} {\bibfnamefont {H.}~\bibnamefont {{Homann}}},
  \bibinfo {author} {\bibfnamefont {D.}~\bibnamefont {{Lamb}}}, \bibinfo
  {author} {\bibfnamefont {A.~S.}\ \bibnamefont {{Lanotte}}}, \bibinfo {author}
  {\bibfnamefont {E.}~\bibnamefont {{L{\'e}v{\`e}que}}}, \bibinfo {author}
  {\bibfnamefont {B.}~\bibnamefont {{L{\"u}thi}}}, \bibinfo {author}
  {\bibfnamefont {J.}~\bibnamefont {{Mann}}}, \bibinfo {author} {\bibfnamefont
  {N.}~\bibnamefont {{Mordant}}}, \bibinfo {author} {\bibfnamefont {W.-C.}\
  \bibnamefont {{M{\"u}ller}}}, \bibinfo {author} {\bibfnamefont
  {S.}~\bibnamefont {{Ott}}}, \bibinfo {author} {\bibfnamefont {N.~T.}\
  \bibnamefont {{Ouellette}}}, \bibinfo {author} {\bibfnamefont {J.-F.}\
  \bibnamefont {{Pinton}}}, \bibinfo {author} {\bibfnamefont {S.~B.}\
  \bibnamefont {{Pope}}}, \bibinfo {author} {\bibfnamefont {S.~G.}\
  \bibnamefont {{Roux}}}, \bibinfo {author} {\bibfnamefont {F.}~\bibnamefont
  {{Toschi}}}, \bibinfo {author} {\bibfnamefont {H.}~\bibnamefont {{Xu}}}, \
  and\ \bibinfo {author} {\bibfnamefont {P.~K.}\ \bibnamefont {{Yeung}}},\
  }\href {\doibase 10.1103/PhysRevLett.100.254504} {\bibfield  {journal}
  {\bibinfo  {journal} {Physical Review Letters}\ }\textbf {\bibinfo {volume}
  {100}},\ \bibinfo {eid} {254504} (\bibinfo {year} {2008})},\ \Eprint
  {http://arxiv.org/abs/0802.3776} {arXiv:0802.3776 [nlin.CD]} \BibitemShut
  {NoStop}%
\bibitem [{\citenamefont {{Benzi}}\ \emph {et~al.}(2009)\citenamefont
  {{Benzi}}, \citenamefont {{Biferale}}, \citenamefont {{Calzavarini}},
  \citenamefont {{Lohse}},\ and\ \citenamefont
  {{Toschi}}}]{2009PhRvE..80f6318B}%
  \BibitemOpen
  \bibfield  {author} {\bibinfo {author} {\bibfnamefont {R.}~\bibnamefont
  {{Benzi}}}, \bibinfo {author} {\bibfnamefont {L.}~\bibnamefont {{Biferale}}},
  \bibinfo {author} {\bibfnamefont {E.}~\bibnamefont {{Calzavarini}}}, \bibinfo
  {author} {\bibfnamefont {D.}~\bibnamefont {{Lohse}}}, \ and\ \bibinfo
  {author} {\bibfnamefont {F.}~\bibnamefont {{Toschi}}},\ }\href {\doibase
  10.1103/PhysRevE.80.066318} {\bibfield  {journal} {\bibinfo  {journal}
  {\pre}\ }\textbf {\bibinfo {volume} {80}},\ \bibinfo {eid} {066318} (\bibinfo
  {year} {2009})},\ \Eprint {http://arxiv.org/abs/0806.4762} {arXiv:0806.4762
  [physics.flu-dyn]} \BibitemShut {NoStop}%
\bibitem [{\citenamefont {{Bec}}\ \emph
  {et~al.}(2010{\natexlab{b}})\citenamefont {{Bec}}, \citenamefont
  {{Biferale}}, \citenamefont {{Lanotte}}, \citenamefont {{Scagliarini}},\ and\
  \citenamefont {{Toschi}}}]{2010JFM...645..497B}%
  \BibitemOpen
  \bibfield  {author} {\bibinfo {author} {\bibfnamefont {J.}~\bibnamefont
  {{Bec}}}, \bibinfo {author} {\bibfnamefont {L.}~\bibnamefont {{Biferale}}},
  \bibinfo {author} {\bibfnamefont {A.~S.}\ \bibnamefont {{Lanotte}}}, \bibinfo
  {author} {\bibfnamefont {A.}~\bibnamefont {{Scagliarini}}}, \ and\ \bibinfo
  {author} {\bibfnamefont {F.}~\bibnamefont {{Toschi}}},\ }\href {\doibase
  10.1017/S0022112009992783} {\bibfield  {journal} {\bibinfo  {journal}
  {Journal of Fluid Mechanics}\ }\textbf {\bibinfo {volume} {645}},\ \bibinfo
  {pages} {497} (\bibinfo {year} {2010}{\natexlab{b}})},\ \Eprint
  {http://arxiv.org/abs/0904.2314} {arXiv:0904.2314 [physics.flu-dyn]}
  \BibitemShut {NoStop}%
\bibitem [{\citenamefont {{Lanotte}}\ \emph {et~al.}(2011)\citenamefont
  {{Lanotte}}, \citenamefont {{Calzavarini}}, \citenamefont {{Toschi}},
  \citenamefont {{Bec}}, \citenamefont {{Biferale}},\ and\ \citenamefont
  {{Cencini}}}]{RM-2007-GRAD-2048}%
  \BibitemOpen
  \bibfield  {author} {\bibinfo {author} {\bibfnamefont {A.~S.}\ \bibnamefont
  {{Lanotte}}}, \bibinfo {author} {\bibfnamefont {E.}~\bibnamefont
  {{Calzavarini}}}, \bibinfo {author} {\bibfnamefont {F.}~\bibnamefont
  {{Toschi}}}, \bibinfo {author} {\bibfnamefont {J.}~\bibnamefont {{Bec}}},
  \bibinfo {author} {\bibfnamefont {L.}~\bibnamefont {{Biferale}}}, \ and\
  \bibinfo {author} {\bibfnamefont {M.}~\bibnamefont {{Cencini}}},\ }\href@noop
  {} {} (\bibinfo {year} {2011}),\ \bibinfo {note}
  {\url{http://data.3tu.nl/repository/uuid:f7cd7b9d-ae4e-498e-92b4-7efe2d350d86}
  [Online; accessed 16-Dec-2013 and 24-Jan-2014]}\BibitemShut {NoStop}%
\bibitem [{\citenamefont {{Kolmogorov}}(1962)}]{1962JFM....13...82K}%
  \BibitemOpen
  \bibfield  {author} {\bibinfo {author} {\bibfnamefont {A.~N.}\ \bibnamefont
  {{Kolmogorov}}},\ }\href {\doibase 10.1017/S0022112062000518} {\bibfield
  {journal} {\bibinfo  {journal} {Journal of Fluid Mechanics}\ }\textbf
  {\bibinfo {volume} {13}},\ \bibinfo {pages} {82} (\bibinfo {year}
  {1962})}\BibitemShut {NoStop}%
\bibitem [{Note2()}]{Note2}%
  \BibitemOpen
  \bibinfo {note} {They argue that the scaling should change to $r^{-4/3}$ for
  very high Reynolds numbers $Re_\lambda \ge 600$}\BibitemShut {NoStop}%
\bibitem [{\citenamefont {{Meneveau}}\ \emph {et~al.}(1990)\citenamefont
  {{Meneveau}}, \citenamefont {{Sreenivasan}}, \citenamefont {{Kailasnath}},\
  and\ \citenamefont {{Fan}}}]{1990PhRvA..41..894M}%
  \BibitemOpen
  \bibfield  {author} {\bibinfo {author} {\bibfnamefont {C.}~\bibnamefont
  {{Meneveau}}}, \bibinfo {author} {\bibfnamefont {K.~R.}\ \bibnamefont
  {{Sreenivasan}}}, \bibinfo {author} {\bibfnamefont {P.}~\bibnamefont
  {{Kailasnath}}}, \ and\ \bibinfo {author} {\bibfnamefont {M.~S.}\
  \bibnamefont {{Fan}}},\ }\href {\doibase 10.1103/PhysRevA.41.894} {\bibfield
  {journal} {\bibinfo  {journal} {\pra}\ }\textbf {\bibinfo {volume} {41}},\
  \bibinfo {pages} {894} (\bibinfo {year} {1990})}\BibitemShut {NoStop}%
\bibitem [{\citenamefont {{Cao}}\ \emph {et~al.}(1996)\citenamefont {{Cao}},
  \citenamefont {{Chen}},\ and\ \citenamefont
  {{Sreenivasan}}}]{1996PhRvL..77.3799C}%
  \BibitemOpen
  \bibfield  {author} {\bibinfo {author} {\bibfnamefont {N.}~\bibnamefont
  {{Cao}}}, \bibinfo {author} {\bibfnamefont {S.}~\bibnamefont {{Chen}}}, \
  and\ \bibinfo {author} {\bibfnamefont {K.~R.}\ \bibnamefont
  {{Sreenivasan}}},\ }\href {\doibase 10.1103/PhysRevLett.77.3799} {\bibfield
  {journal} {\bibinfo  {journal} {Physical Review Letters}\ }\textbf {\bibinfo
  {volume} {77}},\ \bibinfo {pages} {3799} (\bibinfo {year}
  {1996})}\BibitemShut {NoStop}%
\bibitem [{\citenamefont {{Chen}}\ \emph {et~al.}(1997)\citenamefont {{Chen}},
  \citenamefont {{Sreenivasan}},\ and\ \citenamefont
  {{Nelkin}}}]{1997PhRvL..79.1253C}%
  \BibitemOpen
  \bibfield  {author} {\bibinfo {author} {\bibfnamefont {S.}~\bibnamefont
  {{Chen}}}, \bibinfo {author} {\bibfnamefont {K.~R.}\ \bibnamefont
  {{Sreenivasan}}}, \ and\ \bibinfo {author} {\bibfnamefont {M.}~\bibnamefont
  {{Nelkin}}},\ }\href {\doibase 10.1103/PhysRevLett.79.1253} {\bibfield
  {journal} {\bibinfo  {journal} {Physical Review Letters}\ }\textbf {\bibinfo
  {volume} {79}},\ \bibinfo {pages} {1253} (\bibinfo {year}
  {1997})}\BibitemShut {NoStop}%
\bibitem [{\citenamefont {{Sreenivasan}}\ and\ \citenamefont
  {{Antonia}}(1997)}]{1997AnRFM..29..435S}%
  \BibitemOpen
  \bibfield  {author} {\bibinfo {author} {\bibfnamefont {K.~R.}\ \bibnamefont
  {{Sreenivasan}}}\ and\ \bibinfo {author} {\bibfnamefont {R.~A.}\ \bibnamefont
  {{Antonia}}},\ }\href {\doibase 10.1146/annurev.fluid.29.1.435} {\bibfield
  {journal} {\bibinfo  {journal} {Annual Review of Fluid Mechanics}\ }\textbf
  {\bibinfo {volume} {29}},\ \bibinfo {pages} {435} (\bibinfo {year}
  {1997})}\BibitemShut {NoStop}%
\bibitem [{\citenamefont {{Chun}}\ \emph {et~al.}(2005)\citenamefont {{Chun}},
  \citenamefont {{Koch}}, \citenamefont {{Rani}}, \citenamefont {{Ahluwalia}},\
  and\ \citenamefont {{Collins}}}]{Chunetal2005}%
  \BibitemOpen
  \bibfield  {author} {\bibinfo {author} {\bibfnamefont {J.}~\bibnamefont
  {{Chun}}}, \bibinfo {author} {\bibfnamefont {D.~L.}\ \bibnamefont {{Koch}}},
  \bibinfo {author} {\bibfnamefont {S.~L.}\ \bibnamefont {{Rani}}}, \bibinfo
  {author} {\bibfnamefont {A.}~\bibnamefont {{Ahluwalia}}}, \ and\ \bibinfo
  {author} {\bibfnamefont {L.~R.}\ \bibnamefont {{Collins}}},\ }\href {\doibase
  10.1017/S0022112005004568} {\bibfield  {journal} {\bibinfo  {journal}
  {Journal of Fluid Mechanics}\ }\textbf {\bibinfo {volume} {536}},\ \bibinfo
  {pages} {219} (\bibinfo {year} {2005})}\BibitemShut {NoStop}%
\bibitem [{\citenamefont {Meek}\ and\ \citenamefont
  {Jones}(1973)}]{Meek:1973qv}%
  \BibitemOpen
  \bibfield  {author} {\bibinfo {author} {\bibfnamefont {C.~C.}\ \bibnamefont
  {Meek}}\ and\ \bibinfo {author} {\bibfnamefont {B.~G.}\ \bibnamefont
  {Jones}},\ }\bibfield  {booktitle} {\emph {\bibinfo {booktitle} {Journal of
  the Atmospheric Sciences}},\ }\href {\doibase
  10.1175/1520-0469(1973)030<0239:SOTBOH>2.0.CO;2} {\bibfield  {journal}
  {\bibinfo  {journal} {Journal of the Atmospheric Sciences}\ }\textbf
  {\bibinfo {volume} {30}},\ \bibinfo {pages} {239} (\bibinfo {year}
  {1973})}\BibitemShut {NoStop}%
\bibitem [{\citenamefont {{V\"{o}lk}}\ \emph {et~al.}(1980)\citenamefont
  {{V\"{o}lk}}, \citenamefont {{Jones}}, \citenamefont {{Morfill}},\ and\
  \citenamefont {{Roeser}}}]{1980A&A....85..316V}%
  \BibitemOpen
  \bibfield  {author} {\bibinfo {author} {\bibfnamefont {H.~J.}\ \bibnamefont
  {{V\"{o}lk}}}, \bibinfo {author} {\bibfnamefont {F.~C.}\ \bibnamefont
  {{Jones}}}, \bibinfo {author} {\bibfnamefont {G.~E.}\ \bibnamefont
  {{Morfill}}}, \ and\ \bibinfo {author} {\bibfnamefont {S.}~\bibnamefont
  {{Roeser}}},\ }\href@noop {} {\bibfield  {journal} {\bibinfo  {journal}
  {\aap}\ }\textbf {\bibinfo {volume} {85}},\ \bibinfo {pages} {316} (\bibinfo
  {year} {1980})}\BibitemShut {NoStop}%
\bibitem [{\citenamefont {{Markiewicz}}\ \emph {et~al.}(1991)\citenamefont
  {{Markiewicz}}, \citenamefont {{Mizuno}},\ and\ \citenamefont
  {{Voelk}}}]{1991A&A...242..286M}%
  \BibitemOpen
  \bibfield  {author} {\bibinfo {author} {\bibfnamefont {W.~J.}\ \bibnamefont
  {{Markiewicz}}}, \bibinfo {author} {\bibfnamefont {H.}~\bibnamefont
  {{Mizuno}}}, \ and\ \bibinfo {author} {\bibfnamefont {H.~J.}\ \bibnamefont
  {{Voelk}}},\ }\href@noop {} {\bibfield  {journal} {\bibinfo  {journal}
  {\aap}\ }\textbf {\bibinfo {volume} {242}},\ \bibinfo {pages} {286} (\bibinfo
  {year} {1991})}\BibitemShut {NoStop}%
\bibitem [{\citenamefont {{Cuzzi}}\ and\ \citenamefont
  {{Hogan}}(2003)}]{2003Icar..164..127C}%
  \BibitemOpen
  \bibfield  {author} {\bibinfo {author} {\bibfnamefont {J.~N.}\ \bibnamefont
  {{Cuzzi}}}\ and\ \bibinfo {author} {\bibfnamefont {R.~C.}\ \bibnamefont
  {{Hogan}}},\ }\href {\doibase 10.1016/S0019-1035(03)00104-0} {\bibfield
  {journal} {\bibinfo  {journal} {\icarus}\ }\textbf {\bibinfo {volume}
  {164}},\ \bibinfo {pages} {127} (\bibinfo {year} {2003})}\BibitemShut
  {NoStop}%
\bibitem [{\citenamefont {{Ormel}}\ and\ \citenamefont
  {{Cuzzi}}(2007)}]{2007A&A...466..413O}%
  \BibitemOpen
  \bibfield  {author} {\bibinfo {author} {\bibfnamefont {C.~W.}\ \bibnamefont
  {{Ormel}}}\ and\ \bibinfo {author} {\bibfnamefont {J.~N.}\ \bibnamefont
  {{Cuzzi}}},\ }\href {\doibase 10.1051/0004-6361:20066899} {\bibfield
  {journal} {\bibinfo  {journal} {\aap}\ }\textbf {\bibinfo {volume} {466}},\
  \bibinfo {pages} {413} (\bibinfo {year} {2007})},\ \Eprint
  {http://arxiv.org/abs/astro-ph/0702303} {astro-ph/0702303} \BibitemShut
  {NoStop}%
\bibitem [{\citenamefont {{Hubbard}}(2012)}]{2012MNRAS.426..784H}%
  \BibitemOpen
  \bibfield  {author} {\bibinfo {author} {\bibfnamefont {A.}~\bibnamefont
  {{Hubbard}}},\ }\href {\doibase 10.1111/j.1365-2966.2012.21758.x} {\bibfield
  {journal} {\bibinfo  {journal} {\mnras}\ }\textbf {\bibinfo {volume} {426}},\
  \bibinfo {pages} {784} (\bibinfo {year} {2012})},\ \Eprint
  {http://arxiv.org/abs/1207.5365} {arXiv:1207.5365 [astro-ph.EP]} \BibitemShut
  {NoStop}%
\bibitem [{\citenamefont {{Bec}}\ and\ \citenamefont
  {{Ch{\'e}trite}}(2007)}]{BecChetrite2007}%
  \BibitemOpen
  \bibfield  {author} {\bibinfo {author} {\bibfnamefont {J.}~\bibnamefont
  {{Bec}}}\ and\ \bibinfo {author} {\bibfnamefont {R.}~\bibnamefont
  {{Ch{\'e}trite}}},\ }\href {\doibase 10.1088/1367-2630/9/3/077} {\bibfield
  {journal} {\bibinfo  {journal} {New Journal of Physics}\ }\textbf {\bibinfo
  {volume} {9}},\ \bibinfo {pages} {77} (\bibinfo {year} {2007})},\ \Eprint
  {http://arxiv.org/abs/nlin/0701033} {nlin/0701033} \BibitemShut {NoStop}%
\bibitem [{\citenamefont {{Hogan}}\ \emph {et~al.}(1999)\citenamefont
  {{Hogan}}, \citenamefont {{Cuzzi}},\ and\ \citenamefont
  {{Dobrovolskis}}}]{Hoganetal1999}%
  \BibitemOpen
  \bibfield  {author} {\bibinfo {author} {\bibfnamefont {R.~C.}\ \bibnamefont
  {{Hogan}}}, \bibinfo {author} {\bibfnamefont {J.~N.}\ \bibnamefont
  {{Cuzzi}}}, \ and\ \bibinfo {author} {\bibfnamefont {A.~R.}\ \bibnamefont
  {{Dobrovolskis}}},\ }\href {\doibase 10.1103/PhysRevE.60.1674} {\bibfield
  {journal} {\bibinfo  {journal} {\pre}\ }\textbf {\bibinfo {volume} {60}},\
  \bibinfo {pages} {1674} (\bibinfo {year} {1999})}\BibitemShut {NoStop}%
\bibitem [{\citenamefont {{Bec}}(2003)}]{Bec2003}%
  \BibitemOpen
  \bibfield  {author} {\bibinfo {author} {\bibfnamefont {J.}~\bibnamefont
  {{Bec}}},\ }\href {\doibase 10.1063/1.1612500} {\bibfield  {journal}
  {\bibinfo  {journal} {Physics of Fluids}\ }\textbf {\bibinfo {volume} {15}},\
  \bibinfo {pages} {L81} (\bibinfo {year} {2003})},\ \Eprint
  {http://arxiv.org/abs/nlin/0306049} {nlin/0306049} \BibitemShut {NoStop}%
\bibitem [{\citenamefont {{Bec}}\ \emph {et~al.}(2006)\citenamefont {{Bec}},
  \citenamefont {{Biferale}}, \citenamefont {{Boffetta}}, \citenamefont
  {{Cencini}}, \citenamefont {{Musacchio}},\ and\ \citenamefont
  {{Toschi}}}]{Becetal2006}%
  \BibitemOpen
  \bibfield  {author} {\bibinfo {author} {\bibfnamefont {J.}~\bibnamefont
  {{Bec}}}, \bibinfo {author} {\bibfnamefont {L.}~\bibnamefont {{Biferale}}},
  \bibinfo {author} {\bibfnamefont {G.}~\bibnamefont {{Boffetta}}}, \bibinfo
  {author} {\bibfnamefont {M.}~\bibnamefont {{Cencini}}}, \bibinfo {author}
  {\bibfnamefont {S.}~\bibnamefont {{Musacchio}}}, \ and\ \bibinfo {author}
  {\bibfnamefont {F.}~\bibnamefont {{Toschi}}},\ }\href {\doibase
  10.1063/1.2349587} {\bibfield  {journal} {\bibinfo  {journal} {Physics of
  Fluids}\ }\textbf {\bibinfo {volume} {18}},\ \bibinfo {pages} {091702}
  (\bibinfo {year} {2006})},\ \Eprint {http://arxiv.org/abs/nlin/0606024}
  {nlin/0606024} \BibitemShut {NoStop}%
\bibitem [{\citenamefont {{Chhabra}}\ and\ \citenamefont
  {{Jensen}}(1989)}]{1989PhRvL..62.1327C}%
  \BibitemOpen
  \bibfield  {author} {\bibinfo {author} {\bibfnamefont {A.}~\bibnamefont
  {{Chhabra}}}\ and\ \bibinfo {author} {\bibfnamefont {R.~V.}\ \bibnamefont
  {{Jensen}}},\ }\href {\doibase 10.1103/PhysRevLett.62.1327} {\bibfield
  {journal} {\bibinfo  {journal} {Physical Review Letters}\ }\textbf {\bibinfo
  {volume} {62}},\ \bibinfo {pages} {1327} (\bibinfo {year}
  {1989})}\BibitemShut {NoStop}%
\end{thebibliography}%

\end{document}